\documentclass[twocolumn,twocolappendix]{aastex63}
\usepackage{fp}
\usepackage{bm}

\shorttitle{Superphot}
\shortauthors{Hosseinzadeh et al.}

\newcommand{\OKC}{\affiliation{Oskar Klein Centre, Department of Astronomy, Stockholm University, Albanova University Centre, SE-106 91 Stockholm, Sweden}}

\newcommand{\STScI}{\affiliation{Space Telescope Science Institute, 3700 San Martin Drive, Baltimore, MD 21218-2410, USA}}

\newcommand{\CfA}{\affiliation{Center for Astrophysics \textbar{} Harvard \& Smithsonian, 60 Garden Street, Cambridge, MA 02138-1516, USA}}

\newcommand{\Carnegie}{\affiliation{Observatories of the Carnegie Institute for Science, 813 Santa Barbara Street, Pasadena, CA 91101-1232, USA}}

\newcommand{\CIERA}{\affiliation{Center for Interdisciplinary Exploration and Research in Astrophysics and Department of Physics and Astronomy, \\Northwestern University, 2145 Sheridan Road, Evanston, IL 60208-3112, USA}}

\newcommand{\CMU}{\affiliation{Department of Physics, Carnegie Mellon University, 5000 Forbes Avenue, Pittsburgh, PA 15213-3815, USA}}

\newcommand{\IfA}{\affiliation{Institute for Astronomy, University of Hawai`i, 2680 Woodlawn Drive, Honolulu, HI 96822-1839, USA}}
\newcommand{\UCSC}{\affiliation{Department of Astronomy and Astrophysics, University of California, Santa Cruz, CA 95064-1077, USA}}
\newcommand{\Purdue}{\affiliation{Department of Physics and Astronomy, Purdue University, 525 Northwestern Avenue, West Lafayette, IN 47907-2036, USA}}
\newcommand{\Princeton}{\affiliation{Department of Astrophysical Sciences, Princeton University, 4 Ivy Lane, Princeton, NJ 08540-7219, USA}}
\newcommand{\Moore}{\affiliation{Gordon and Betty Moore Foundation, 1661 Page Mill Road, Palo Alto, CA 94304-1209, USA}}
\newcommand{\Durham}{\affiliation{Department of Physics, Durham University, South Road, Durham, DH1 3LE, UK}}
\newcommand{\JHU}{\affiliation{Department of Physics and Astronomy, The Johns Hopkins University, 3400 North Charles Street, Baltimore, MD 21218, USA}}
\newcommand{\Toronto}{\affiliation{David A.\ Dunlap Department of Astronomy and Astrophysics, University of Toronto,\\ 50 St.\ George Street, Toronto, Ontario, M5S 3H4 Canada}}
\newcommand{\Duke}{\affiliation{Department of Physics, Duke University, Campus Box 90305, Durham, NC 27708, USA}}

\newcommand{\Columbia}{\affiliation{Simons Junior Fellow, Department of Astronomy, Columbia University, New York, NY 10027-6601, USA}}
\newcommand{\NCU}{\affiliation{Graduate Institute of Astronomy, National Central University, 300 Jhongda Road, Zhongli, Taoyuan, 32001, Taiwan}}

\begin{document}

\newcommand{\Ntotal}{5243}

\newcommand{\Nhostspec}{4233}
\newcommand{\NhostspecUnclassified}{3600}
\newcommand{\NLowRcc}{1177}
\newcommand{\NKindaLowRcc}{174}
\newcommand{\NQSO}{199}

\newcommand{\specSLSN}{17}
\newcommand{\specII}{93}
\newcommand{\specIIn}{24}
\newcommand{\specIa}{404}
\newcommand{\specIbc}{19}
\FPeval{\Ntrain}{clip(\specSLSN+\specII+\specIIn+\specIa+\specIbc)}
\newcommand{\Nrare}{16}
\FPeval{\Nspec}{clip(\Nrare+\Ntrain)}

\newcommand{\photSLSN}{58}
\newcommand{\photII}{521}
\newcommand{\photIIn}{181}
\newcommand{\photIa}{1257}
\newcommand{\photIbc}{298}
\FPeval{\Ntest}{clip(\photSLSN+\photII+\photIIn+\photIa+\photIbc)}

\newcommand{\accuracy}{82\%}

\newcommand{\completenessSLSN}{82\%}
\newcommand{\completenessII}{89\%}
\newcommand{\completenessIIn}{50\%}
\newcommand{\completenessIa}{85\%}
\newcommand{\completenessIbc}{37\%}
\newcommand{\cmIInII}{17\%}
\newcommand{\cmIaIbc}{5\%}
\newcommand{\cmIbcII}{42\%}

\newcommand{\puritySLSN}{67\%}
\newcommand{\purityII}{73\%}
\newcommand{\purityIIn}{39\%}
\newcommand{\purityIa}{96\%}
\newcommand{\purityIbc}{21\%}
\newcommand{\pmIaIbc}{65\%}

\newcommand{\completenessCCSN}{95\%}
\newcommand{\purityCCSN}{66\%}
\newcommand{\completenessIaBinary}{81\%}
\newcommand{\purityIaBinary}{98\%}

\newcommand{\expectedAgreement}{74\%}
\newcommand{\agreeVillar}{74\%}
\newcommand{\agreeVillarTrain}{83\%}
\newcommand{\agreeSLSN}{48\%}
\newcommand{\agreeII}{65\%}
\newcommand{\agreeIIn}{25\%}
\newcommand{\agreeIa}{92\%}
\newcommand{\agreeIbc}{43\%}
\newcommand{\amSLSNIa}{41\%}
\newcommand{\amSLSNIaabs}{24}
\newcommand{\amIbcIaabs}{94}
\newcommand{\agreePSNID}{76\%}
\newcommand{\agreeNN}{83\%}
\newcommand{\agreeFitprob}{68\%}

\newcommand{\confidenceThresh}{$p \geq 0.75$}
\newcommand{\exclusionAtThresh}{39\%}
\newcommand{\exclusionAtThreshCCSN}{48\%}
\newcommand{\completenessAtThresh}{85\%}

\title{Photometric Classification of \Ntest{} Pan-STARRS1 Supernovae with Superphot}

\correspondingauthor{Griffin Hosseinzadeh}
\email{griffin.hosseinzadeh@cfa.harvard.edu}

\author[0000-0002-0832-2974]{Griffin~Hosseinzadeh}
\CfA

\author[0000-0001-5123-6388]{Frederick~Dauphin}
\CfA\CMU\STScI

\author[0000-0002-5814-4061]{V.~Ashley~Villar}
\CfA\Columbia

\author[0000-0002-9392-9681]{Edo~Berger}
\CfA

\author[0000-0002-6230-0151]{David~O.~Jones}
\UCSC

\author{Peter~Challis}
\CfA

\author[0000-0002-7706-5668]{Ryan~Chornock}
\CIERA

\author[0000-0001-7081-0082]{Maria~R.~Drout}
\Toronto\Carnegie

\author{Ryan~J.~Foley}
\UCSC

\author[0000-0002-1966-3942]{Robert~P.~Kirshner}
\Moore\CfA

\author[0000-0001-9454-4639]{Ragnhild~Lunnan}
\OKC

\author[0000-0003-4768-7586]{Raffaella~Margutti}
\CIERA

\author[0000-0002-0763-3885]{Dan~Milisavljevic}
\Purdue

\author[0000-0001-8415-6720]{Yen-Chen~Pan}
\NCU

\author[0000-0002-4410-5387]{Armin~Rest}
\STScI\JHU

\author[0000-0002-4934-5849]{Daniel~M.~Scolnic}
\Duke

\author[0000-0002-7965-2815]{Eugene~Magnier}
\IfA

\author[0000-0001-9034-4402]{Nigel~Metcalfe}
\Durham

\author[0000-0002-1341-0952]{Richard~Wainscoat}
\IfA

\author[0000-0003-1989-4879]{Christopher~Waters}
\Princeton

\begin{abstract}
The classification of supernovae (SNe) and its impact on our understanding of explosion physics and progenitors have traditionally been based on the presence or absence of certain spectral features. However, current and upcoming wide-field time-domain surveys have increased the transient discovery rate far beyond our capacity to obtain even a single spectrum of each new event. We must therefore rely heavily on photometric classification---connecting SN light curves back to their spectroscopically defined classes. Here, we present Superphot, an open-source Python implementation of the machine-learning classification algorithm of \citeauthor{villar_supernova_2019}, and apply it to \Ntest{} previously unclassified transients from the Pan-STARRS1 Medium Deep Survey for which we obtained spectroscopic host-galaxy redshifts. Our classifier achieves an overall accuracy of \accuracy{}, with completenesses and purities of $>$80\% for the best classes (SNe~Ia and superluminous SNe). For the worst performing SN class (SNe~Ibc), the completeness and purity fall to \completenessIbc{} and \purityIbc{}, respectively. Our classifier provides \photIa{} newly classified SNe~Ia, \photII{} SNe~II, \photIbc{} SNe~Ibc, \photIIn{} SNe~IIn, and \photSLSN{} SLSNe. These are among the largest uniformly observed samples of SNe available in the literature and will enable a wide range of statistical studies of each class.
\end{abstract}

\keywords{Supernovae (1668); Astrostatistics (1882); Light curve classification (1954)}

\received{2020 August 11}
\revised{2020 October 2}
\accepted{2020 October 20}
\published{2020 December 17}

\section{Introduction} \label{sec:intro}
Starting with \cite{minkowski_spectra_1941}, supernovae (SNe) have been classified on the basis of their spectra, with hydrogen-poor events being labeled Type~I and hydrogen-rich events Type~II. \cite{uomoto_peculiar_1985} and \cite{wheeler_peculiar_1985} later separated Type~Ia SNe (SNe~Ia), which show silicon absorption (as well as a secondary infrared light-curve peak; \citealt{elias_type_1985}), from the remaining SNe~I, which were later divided into helium-rich SNe~Ib and helium-poor SNe~Ic \citep{wheeler_physical_1986}. Similarly, \cite{schlegel_new_1990} separated SNe~IIn, which show narrow hydrogen emission lines, from the remaining SNe~II. More recently, \cite{quimby_hydrogen-poor_2011} and \cite{gal-yam_luminous_2012} defined a class of superluminous SNe (SLSNe) that are $10\times$--$100\times$ more luminous than the aforementioned classes; though originally identified photometrically, hydrogen-poor SLSNe are now considered a spectroscopic class \citep{quimby_spectra_2018,gal-yam_most_2019}. Subsequent authors have further subdivided all of the aforementioned classes (see \citealt{gal-yam_observational_2016} for a review).

With the advent of high-\'etendue time-domain facilities like the Panoramic Survey Telescope and Rapid Response System 1 (Pan-STARRS1; \citealt{chambers_pan-starrs1_2016}) and Zwicky Transient Facility \citep{bellm_zwicky_2019}, as well as the upcoming Vera C.\ Rubin Observatory \citep{ivezic_lsst_2019}, the transient discovery rate has far exceeded the worldwide capacity for spectroscopic classification. We must therefore rely on photometric classification methods, despite the fact that the classes are defined spectroscopically.

Many previous attempts at photometric classification have focused only on separating SNe~Ia from all other SN classes, for the purpose of measuring cosmological parameters \citep{riess_identification_2004,riess_type_2004,moller_photometric_2016,kimura_single-epoch_2017}. This is a somewhat easier problem because of the relative photometric uniformity of SNe~Ia. Other attempts have relied on simulated data for their training sets \citep{richards_semi-supervised_2012,charnock_deep_2017,kimura_single-epoch_2017,boone_avocado_2019,ishida_optimizing_2019,muthukrishna_rapid_2019}. The use of simulated light curves implies that we understand the full diversity of explosive transients, although it is a logical way forward in the absence of large, uniformly observed photometric data sets.

\defcitealias{villar_supernova_2019}{V19}

\citet[hereafter \citetalias{villar_supernova_2019}]{villar_supernova_2019} presented a method for photometric classification that consists of (1) fitting a highly flexible analytical model to the observed light curve, (2) extracting features from that model light curve, and (3) using supervised machine learning to classify the SN based on those features. \citetalias{villar_supernova_2019} tested 24 different pipelines---consisting of 4 different methods of feature extraction, 2 different methods for balancing the classes in the training set, and 3 different machine-learning algorithms---that they trained on the spectroscopically classified transients in the Pan-STARRS1 Medium Deep Survey (PS1-MDS; \citealt{chambers_pan-starrs1_2016}). This is the largest real data set that has been used to train a multiclass classifier.

Here, we present an open-source Python implementation of this method and apply it to \Ntest{} previously unclassified transients from the same survey for which we obtained spectroscopic host-galaxy redshifts. The code is available on GitHub and Zenodo \citep{hosseinzadeh_superphot_2020} and listed on the Python Package Index under the name Superphot.\footnote{Documentation: \url{https://griffin-h.github.io/superphot/}}\textsuperscript{,}\footnote{``Superphot'' is a blend \citep{algeo_blends_1977} of the words ``supernova photometry'' but is also intended to sound like the SN spectrum-fitting code Superfit \citep{howell_type_2006} used for spectroscopic classification.}

For the purposes of this work, we restrict our classification to the following five broad labels, adopting only slight refinements to the definitions discussed above that are already in common use:
\begin{enumerate}
    \item ``SN~Ia'' refers to hydrogen-poor SNe that show silicon absorption and a secondary infrared light-curve peak.
    \item ``SN~Ibc'' refers to all normal-luminosity, hydrogen-poor SNe that are not SNe~Ia. Due to the small number of these in our training sample, we do not distinguish between SNe~Ib and SNe~Ic.
    \item ``SN~IIn'' refers to SNe that show narrow hydrogen emission lines in their spectra throughout their evolution.
    \item ``SN~II'' refers to all hydrogen-rich SNe that are not SNe~IIn. We do not distinguish between SNe~IIL and SNe~IIP, because it is not clear that these populations are separate \citep{sanders_toward_2015,valenti_diversity_2016}.
    \item ``SLSN'' refers only to hydrogen-poor SLSNe. Hydrogen-rich SLSNe are included in SNe~IIn.
\end{enumerate}

In Section~\ref{sec:data}, we briefly present the photometric data set we use for training and classification, as well as the spectroscopic data set used to determine the host-galaxy redshifts. In Section~\ref{sec:algorithm}, we describe the implementation of the algorithm. In Section ~\ref{sec:validate}, we apply our algorithm to the unclassified light curves and assess its performance. In Section~\ref{sec:discuss}, we discuss its utility for current and future time-domain surveys. In Section~\ref{sec:conclusion}, we conclude with some lessons learned from this case study in photometric classification.

\begin{deluxetable*}{cccCcccCcc}
\tablecaption{Host-galaxy Redshifts from RVSAO\label{tab:rvsao}}
\tablehead{\colhead{Transient} & \colhead{Spec.} & \colhead{Host} & \colhead{Host} & \colhead{MJD of} & \colhead{} & \colhead{Final} & \colhead{RVSAO} & \colhead{Template} & \colhead{Maximum} \\[-6pt]
\colhead{Name} & \colhead{Class.} & \colhead{R.A. (deg)} & \colhead{Decl. (deg)} & \colhead{Observation} & \colhead{Telescope} & \colhead{Redshift} & \colhead{Redshift} & \colhead{Matches} & \colhead{$R_\mathrm{CC}$}}
\startdata
PSc000006 & SN~Ia & 53.3663 & -28.3715 & 57013 & AAT & 0.231 & 0.230 & 8 & 6.56 \\
 &  &  &  &  &  &  & 0.109 & 2 & 2.40 \\
PSc000010 & SN~Ia & 149.7495 & 3.1576 & 57133 & MMT & 0.245 & 0.244 & 13 & 15.0 \\
PSc000011 & SN~Ia & 149.9760 & 2.4106 & 57013 & AAT & 0.380 & 0.731 & 3 & 6.77 \\
 &  &  &  &  &  &  & 0.361 & 2 & 2.58 \\
PSc000012 & \nodata & 150.2308 & 1.8451 & 55296 & MMT & 0.623 & 0.179 & 3 & 3.12 \\
 &  &  &  &  &  &  & 0.341 & 2 & 2.67 \\
 &  &  &  &  &  &  & 0.727 & 2 & 1.89 \\
PSc000013 & \nodata & 149.1394 & 1.5447 & 56564 & MMT & 0.372 & 0.372 & 5 & 5.81 \\
 &  &  &  &  &  &  & 0.927 & 2 & 3.60
\enddata
\tablecomments{This table is available in its entirety in machine-readable form.}
\end{deluxetable*}

\section{Data Set}\label{sec:data}
Pan-STARRS1 is a 1.8~m telescope near the summit of Haleakal\=a, Hawai`i, equipped with a 1.4 gigapixel camera with a 7~deg$^2$ field of view \citep{chambers_pan-starrs1_2016}. PS1-MDS ran from 2009 July to 2014 July using 25\% of the observing time on Pan-STARRS1 and consisted of 10 deep-drilling fields with a three-day cadence in any of five bands: \textit{grizy} \citep{chambers_pan-starrs1_2016}. Images were processed and calibrated using the Image Processing Pipeline \citep{magnier_pan-starrs_2019-2,magnier_pan-starrs_2019-1,magnier_pan-starrs_2019,waters_pan-starrs_2019}. With the exception of $y$, PS1-MDS reached typical depths of 23.3~mag per visit \citep{chambers_pan-starrs1_2016}; the $y$ filter has lower throughput and cadence, so we exclude it from our analysis. As such, our data set is very similar to the proposed depth and cadence of the Rubin Observatory Legacy Survey of Space and Time (LSST), but with $\sim$0.1\% of the time-integrated sky coverage \citep{ivezic_lsst_2019}.

\defcitealias{villar_superraenn_2020}{V20}

Over the course of PS1-MDS, we detected a total of \Ntotal{} SN-like transients\footnote{Transients with three $\mathrm{S/N} \ge 4$ photometric observations in any filter and no history of variability \citep{jones_measuring_2018}.} using \texttt{photpipe} \citep{rest_testing_2005,rest_cosmological_2014}. These light curves are presented in a companion paper by \citet[][hereafter \citetalias{villar_superraenn_2020}]{villar_superraenn_2020} and are available on Zenodo \citep{villar_light_2020}. We obtained spectroscopic classifications of \Nspec{} of these transients. The \Ntrain{} SNe belonging to one of the five classes listed in Section~\ref{sec:intro} comprise our ``training set.'' The \specSLSN{} SLSNe in our data set were previously published by \cite{lunnan_hydrogen-poor_2018}, and 76 of the SN~II light curves were analyzed by \cite{sanders_toward_2015}. The remaining \Nrare{} spectroscopically classified transients belong to less common classes. Because these are too few in number to be used as training samples, we exclude them from our analysis except to explore how they are labeled by our classifier (Section~\ref{sec:rare}).

We also obtained host-galaxy spectra for \Nhostspec{} transients, \NhostspecUnclassified{} of which were not classified spectroscopically: 3434 from MMT, 324 from the Anglo-Australian Telescope (AAT), 301 from WIYN, 169 from the Sloan Digital Sky Survey (SDSS; \citealt{ahumada_16th_2020}), and 5 from Apache Point Observatory (APO). These spectra are available on Zenodo \citep{hosseinzadeh_host_2020}. We used RVSAO \citep{kurtz_rvsao_1998} to cross-correlate these spectra with a library of galaxy templates, including 6 built into RVSAO and an additional 10 from SDSS. Table~\ref{tab:rvsao} logs these observations and lists the modes of the resulting redshift distribution from RVSAO, ignoring matches with $z < 0.005$ ($d_L < 20$~Mpc).

We excluded from our analysis \NLowRcc{} spectra where all template matches had a \cite{tonry_survey_1979} cross-correlation score $R_\mathrm{CC} < 4$, meaning that the cross-correlation redshift may not be reliable, and \NKindaLowRcc{} transients with only a single template match in the range $4 < R_\mathrm{CC} < 5$. We then visually inspected approximately 600 host spectra that met one of three criteria: (1) the best (highest $R_\mathrm{CC}$), median, and modal redshifts did not match; (2) no three redshift estimates matched each other; or (3) the best redshift could have been derived from matching a telluric feature or a known instrumental artifact to a feature in the galaxy template. In almost two-thirds of these cases, we were able to either verify the redshift from RVSAO or determine a new redshift manually. We excluded the remaining third from further analysis. In addition to the redshifts from RVSAO, we supplemented our sample with redshifts from publicly available catalogs\footnote{\cite{1992ApJS...78....1D,2001AJ....122..750I,2003astro.ph..6581C,2004ApJS..155..271S,2005A&A...439..845L,2006MNRAS.372..425C,2006AJ....132.2409N,2007A&A...474..473G,2007ApJS..172...70L,2007A&A...467...73T,2008A&A...477..717B,2008MNRAS.387.1323R,2009ApJ...703L.162F,2009MNRAS.399..683J,2009A&A...495...53L,2009ApJS..182..625O,2009ApJ...704L..98S,2009ApJ...696.1195T,2010A&A...512A..12B,2010ApJ...711..928C,2010MNRAS.401.1429D,2010MNRAS.405.2302H,2010MNRAS.401..294S,2011A&A...529A.135R,2012CBET.3274....1C,2012MNRAS.422...25S,2013ApJS..208....5N,drout_rapidly_2014,2014MNRAS.441.1802K,2014ApJ...787..138L,2015ApJ...807..178W,2018ApJ...858...77H,2019ApJ...877...81M,ahumada_16th_2020,2020MNRAS.496...19L}} using a $1''$ matching radius; in the case of a conflict with RVSAO, we manually inspected our spectrum to determine a final redshift. Lastly, two transient spectra (of PSc110446 and PSc130816) yielded redshifts but not confident classifications; we treat these as unclassified transients with known redshifts.

Finally, we excluded \NQSO{} unclassified transients whose light curves are variable across multiple observing seasons, indicating that they are unlikely to be SNe. The remaining \Ntest{} transients comprise our ``test set'' for photometric classification. \cite{jones_measuring_2017} previously used 1020 SNe~Ia from this data set, some of which were photometrically classified by a different method, to constrain cosmological parameters.

\vspace{1in}
\section{Description of the Algorithm}\label{sec:algorithm}
\subsection{Model Fitting}
We use Equation~1 of \citetalias{villar_supernova_2019} to model the single-band flux of a transient with the following form:

\begin{equation}\label{eq:model}
    F(\Delta t) = \frac{A\left[1-\beta \min(\Delta t, \gamma)\right] \exp\left(-\frac{\max(\Delta t, \gamma) - \gamma}{\tau_\mathrm{fall}}\right)}{1 + \exp(-\frac{\Delta t}{\tau_\mathrm{rise}})}
\end{equation}
where $\Delta t \equiv t - t_0$ is time with respect to a reference epoch in the observer frame. This function has six parameters that are not strictly physical, but roughly correspond to an amplitude ($A$), the ``plateau'' slope and duration\footnote{In the parametrization of \citetalias{villar_supernova_2019}, $\beta \to -\frac{\beta}{A}$ and $\gamma \equiv t_1 - t_0$.} ($\beta$ and $\gamma$), the reference epoch with respect to discovery ($t_0$), and the (exponential) rise and decline times ($\tau_\mathrm{rise}$ and $\tau_\mathrm{fall}$).

To obtain not only the best-fit parameters but a quantification of the uncertainties, we use a Markov Chain Monte Carlo (MCMC) routine to fit each of the observed $griz$ light curves for these six parameters plus an additional intrinsic scatter term, which is added in quadrature with the photometric uncertainties. The seven fit parameters and their priors are listed in Table~\ref{tab:params}, where $U(a, b)$ indicates a uniform distribution between $a$ and $b$, $U_{\log}(a, b)$ indicates a log-uniform distribution between $a$ and $b$, and $N(\mu, \sigma^2)$ indicates a Gaussian distribution with mean $\mu$ and variance $\sigma^2$. These are approximately the same priors used by \citetalias{villar_supernova_2019}, with the exception of $A$, for which they used a uniform prior. All parameters apart from $t_0$ are restricted from taking negative values.

\begin{deluxetable}{CccC}
\tabletypesize\footnotesize
\tablecolumns{4}
\tablecaption{Model Parameters\label{tab:params}}
\tablehead{\colhead{} & \colhead{Parameter} & \colhead{Units} & \colhead{Prior}}
\startdata
A & Amplitude & [flux] & U_{\log}(1, 100F_\mathrm{max}^\mathrm{obs}) \\
\beta & Plateau Slope & days$^{-1}$ & U(0, 0.01) \\
\gamma & Plateau Duration & days & \frac{2}{3}N(5, 25) + \frac{1}{3}N(60, 900) \\
t_0 & Reference Epoch & days & U(-50, 300) \\
\tau_\mathrm{rise} & Rise Time & days & U(0.01, 50) \\
\tau_\mathrm{fall} & Fall Time & days & U(1, 300) \\
\nodata & Intrinsic Scatter & [flux] & N(0, 1) \\
\enddata
\end{deluxetable}

\begin{figure}
    \centering
    \includegraphics[width=\columnwidth]{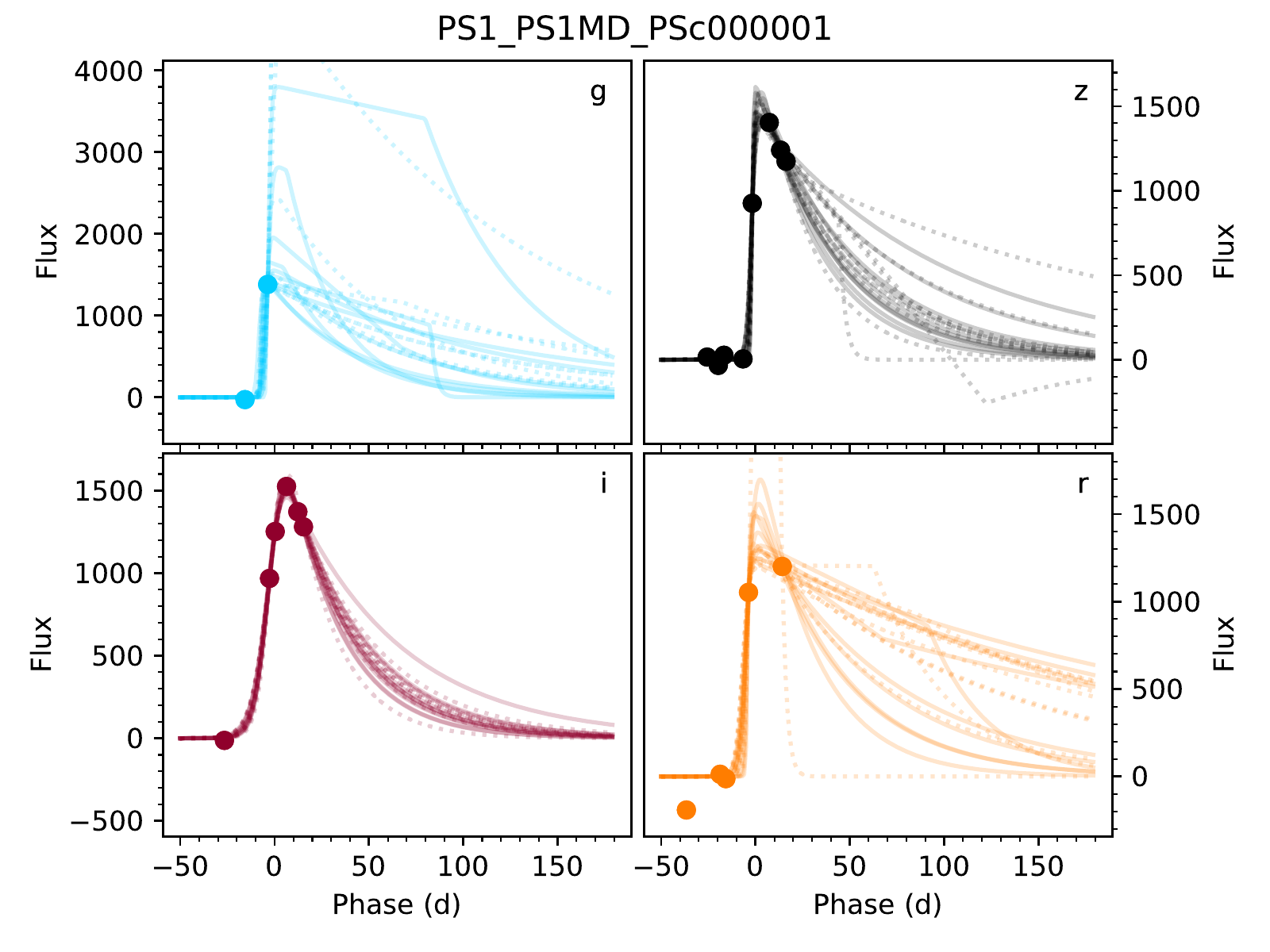}
    \includegraphics[width=\columnwidth]{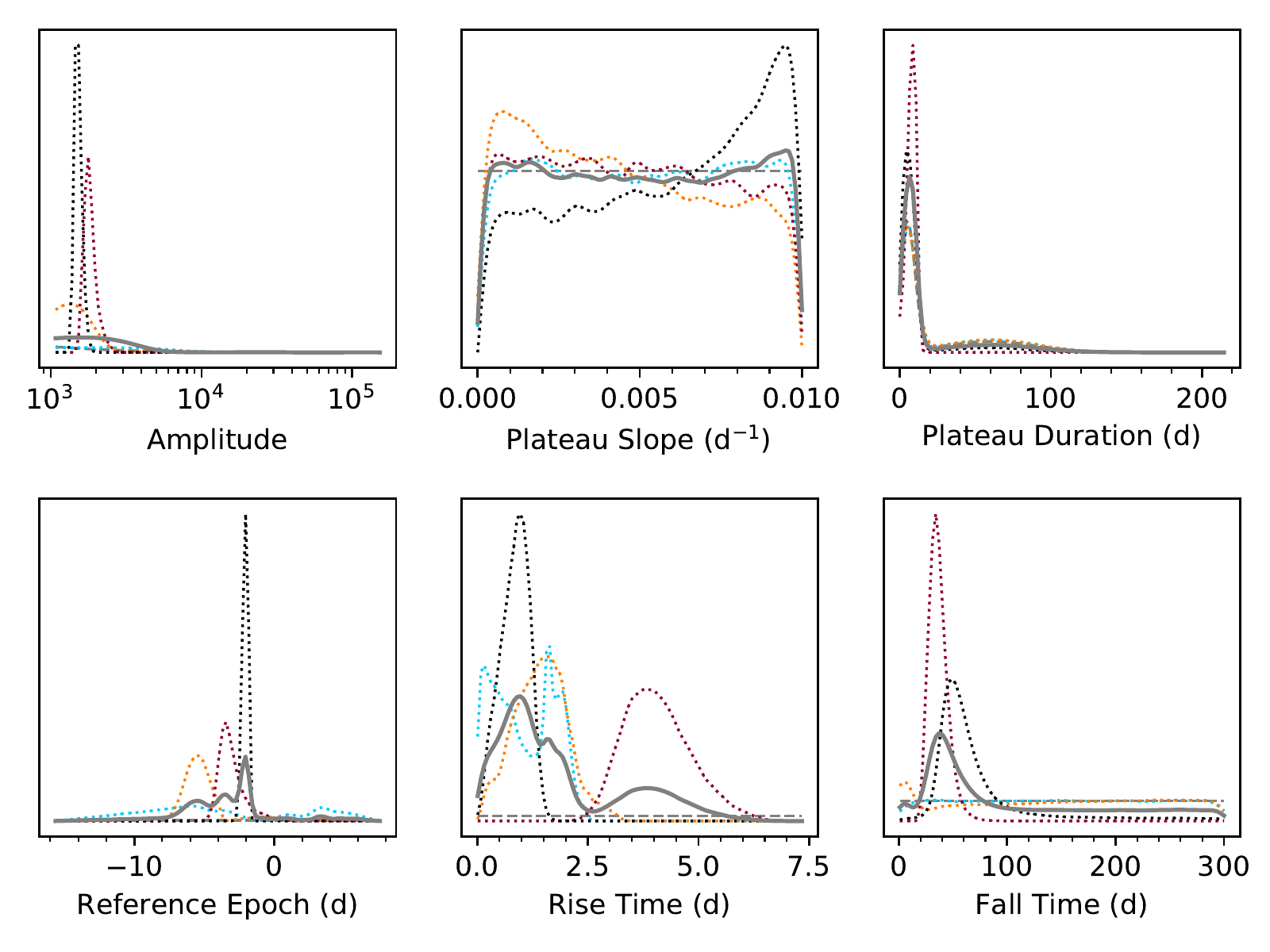}
    \caption{Two-step MCMC fit of Equation~\ref{eq:model} to a sample transient. The top panels show the light-curve models in each of the four filters. Dotted lines represent the results of the first iteration, and solid lines represent the results of the second. The bottom panels show how the first-iteration posteriors are combined to produce the second-iteration priors. The gray dashed line is the first-iteration prior, the colored dotted lines represent the first-iteration posteriors for the four filters (same colors as the upper panels), and the gray solid line is the second-iteration prior.}
    \label{fig:fit}
\end{figure}

Because not all filters ($griz$) are well sampled for all events, we wish to include some cross-filter information in the fit. However, prior to classification, we cannot assume knowledge of the shape and evolution of the spectral energy distribution (SED). We also found it difficult to achieve MCMC convergence within a reasonable time when simultaneously fitting more than 20 parameters. We therefore adopt \citetalias{villar_supernova_2019}'s two-iteration fitting approach that consists of fitting each filter separately, adding the posteriors from those fits together, and using the result as the prior for a second iteration of fitting (see Figure~\ref{fig:fit}). This effectively weights the filters toward being more similar to each other, but without excluding the possibility that they are different. If a given SN was not observed in one or more filters, we average the posterior distributions for all the observed filters and treat that as the posterior for all unobserved filters.

For all fits, we use the Metropolis--Hastings sampling algorithm \citep{metropolis_equation_1953,hastings_monte_1970} in the PyMC3 package \citep{salvatier_probabilistic_2016}. We use 25 chains (walkers), each drawing 25,000 samples for tuning (burn-in) plus 10,000 samples for the posterior. For computational efficiency, we only fit points between $-50$ and +180 days of the discovery date, which encompasses $>$99\% of the $\ge$3$\sigma$ detections in the same observing season. The computations in this paper were run on the FASRC Cannon cluster supported by the FAS Division of Science Research Computing Group at Harvard University.\footnote{The Cannon cluster is named after Annie Jump Cannon, one of the human computers at the Harvard College Observatory and a pioneer in stellar classification.} Tables \ref{tab:results}--\ref{tab:other} list the reduced $\chi^2$ goodness-of-fit statistic (for the model with the median parameters) and the maximum (among the 28 model parameters) \cite{gelman_inference_1992} convergence statistic for each transient. In most cases, both statistics are close to their optimal value of 1. Keep in mind that the reduced $\chi^2$ is infinite or negative when the light curve has 24 or fewer points across the four filters.

\begin{figure}
    \centering
    \includegraphics[width=\columnwidth]{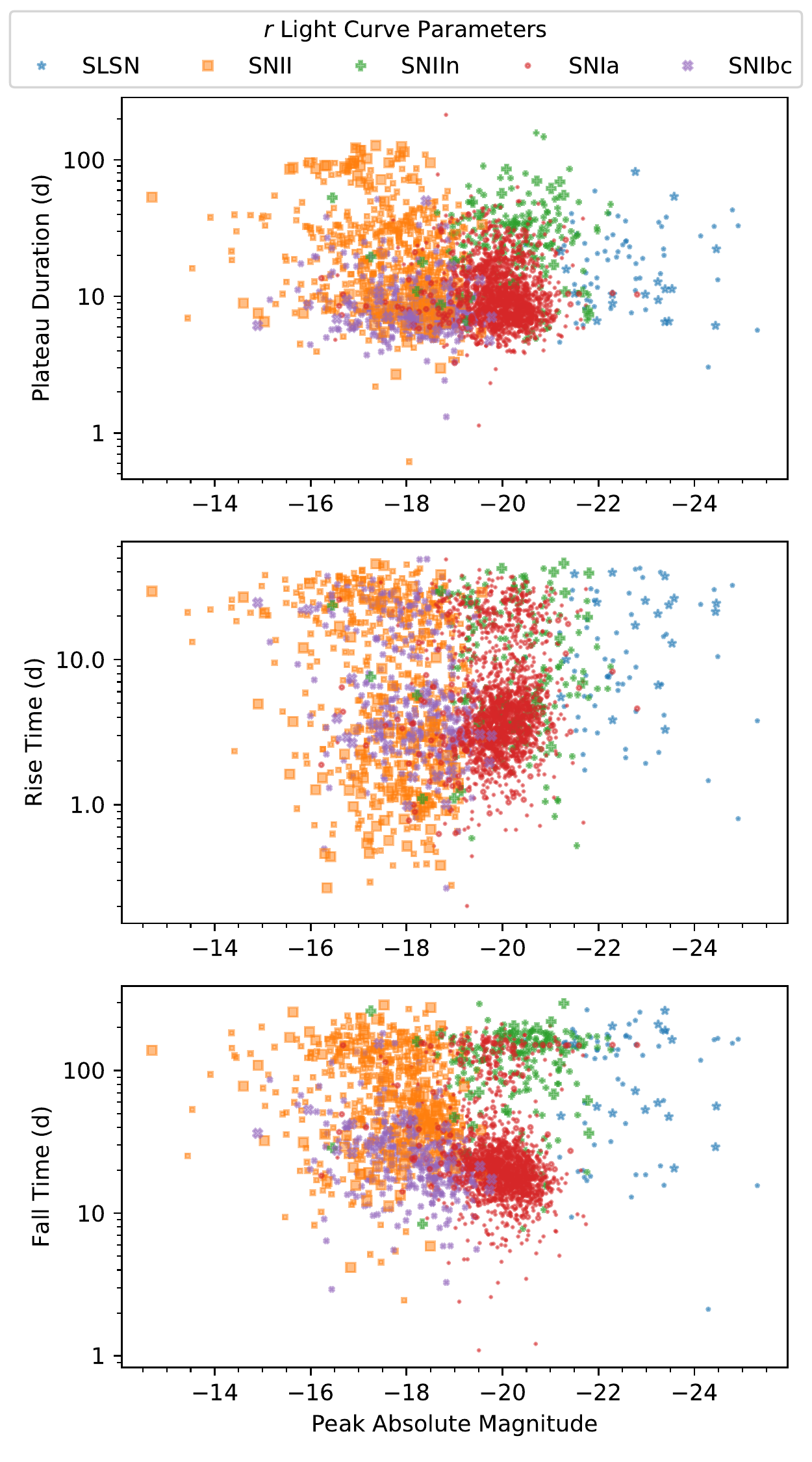}
    \caption{The plateau duration (top), rise time (center), and fall time (bottom) of each model $r$-band light curve plotted against its peak absolute magnitude. Large markers represent the training set and are colored by spectroscopic classification. Small markers represent the test set and are colored by photometric classification. The overlap between the SN classes in this parameter space demonstrates the challenge of photometric classification.}
    \label{fig:params}
\end{figure}

The resulting parameters are shown in Figure~\ref{fig:params}. As expected, the parameters for each photometric class in the test set (small markers) overlap with the parameters for the equivalent spectroscopic class in the training set (large markers). However, there is also significant overlap between different classes in this parameter space. In particular, SNe~Ibc have peak magnitudes and evolution timescales similar to some SNe~II, although SNe~Ibc typically do not have a long plateau. Some SNe~Ia also have long fall times, either because the declining light curve was not well constrained or because the secondary infrared peak was modeled as a smooth decline. We discuss this further in Section~\ref{sec:crossvalidate}.

\subsection{Feature Extraction}
\citetalias{villar_supernova_2019} explored four methods of feature extraction from the model light curves: (1) directly using the model parameters (plus the peak absolute magnitude) as features, (2) hand-selecting features based on the model light curves, (3) performing a principal component analysis (PCA) on the model light curves and using the PCA coefficients (plus the peak absolute magnitude) as features, and (4) using the downsampled model light curves themselves as features. Among their 24 pipelines, there was no clear trend for which of these methods was best. However, as their best-performing pipeline used the PCA method, we adopt that here. Our code also gives the option of using the model parameters directly, but we find that this gives slightly worse results (see Appendix~\ref{sec:hyperparam}).

\begin{figure}
    \centering
    \includegraphics[width=\columnwidth]{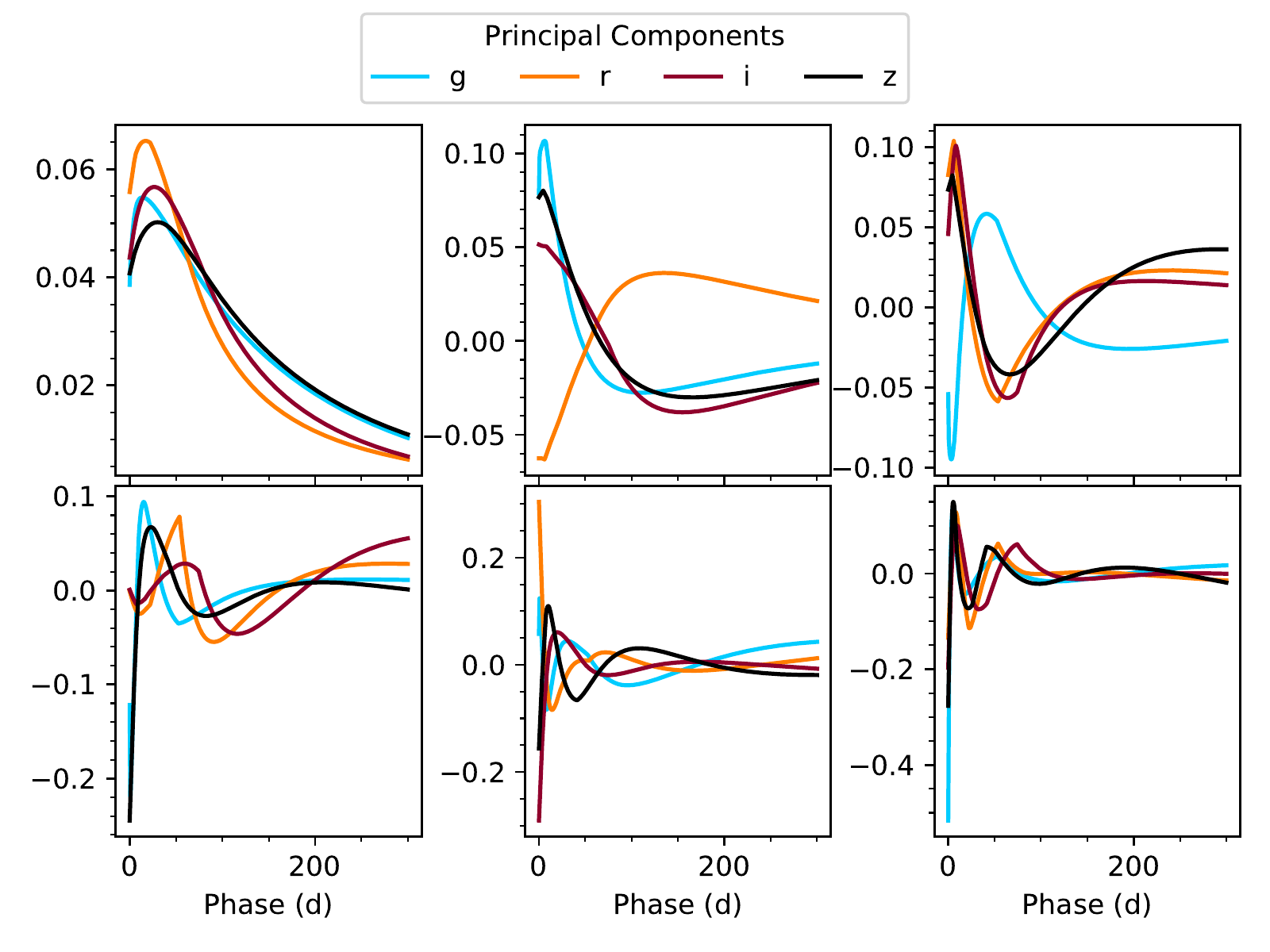}
    \includegraphics[page=1,width=\columnwidth]{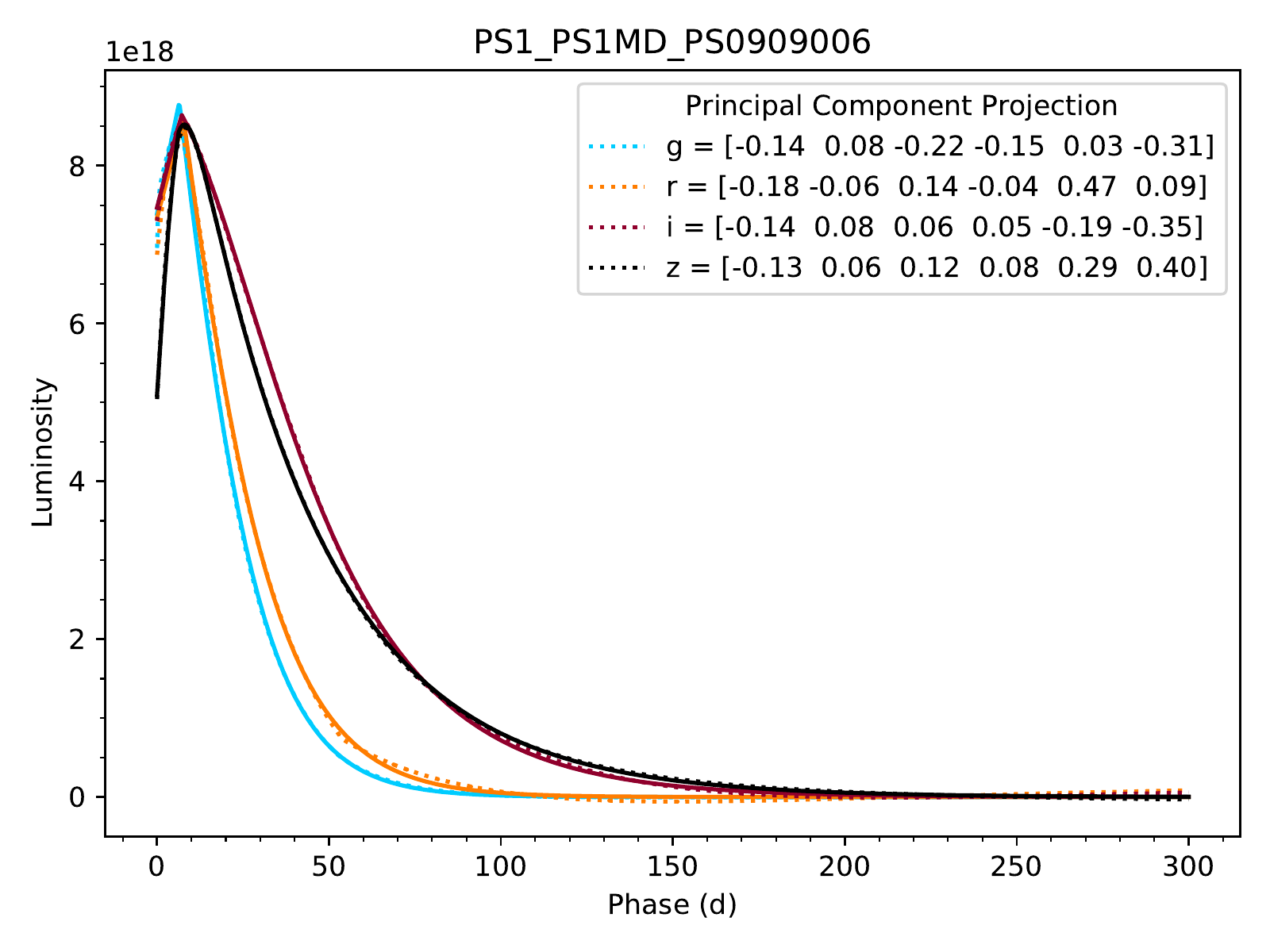}
    \includegraphics[page=17,width=\columnwidth]{train_data_reconstruction.pdf}
    \caption{Top: the top six principal components for light curves in the training set. The projection of light curves onto these axes provides six of the features for each filter, the seventh being peak absolute magnitude. Center: an example of an SN~Ia light curve (solid lines) that is reconstructed nearly perfectly (dotted lines) by its projection onto the principal components. Bottom: an example of an SN~II light curve, which is not reconstructed as well. In both cases, these PCA coefficients were sufficient to classify the SN correctly (see Table~\ref{tab:validation}).}
    \label{fig:pca}
\end{figure}

\defcitealias{planck_collaboration_planck_2016}{Planck Collaboration (2016)}

For the training set, we generate one model light curve for each SN from the median of the posterior parameter distributions. For the validation and test sets, we generate 10 model light curves for each SN by randomly drawing from the posteriors; these uncertainties will be accounted for in our classification probabilities later. We then convert the model fluxes to luminosities. This requires that we know the extinction $E(B-V)$ and redshift $z$ for each transient, where the latter can be measured from the spectrum of either the transient or its host galaxy. In particular, we use luminosity distances calculated with the cosmological parameters of the \citetalias{planck_collaboration_planck_2016}, the Milky Way extinction maps of \cite{schlafly_measuring_2011}, the extinction law of \cite{fitzpatrick_correcting_1999} with $R_V=3.1$, and a cosmological $K$-correction factor of $1+z$ for all filters.\footnote{Correcting to standard rest-frame filters would require detailed knowledge of the SED and time evolution of each transient, which would in turn depend on its classification.} We do not consider host-galaxy extinction, because we have no way of estimating it for transients in the test set. \citetalias{villar_supernova_2019} found that correcting to rest-frame times gave worse results, so we leave all times in the observer frame.

For each filter, we then perform a PCA on the model light curves (in luminosity) in our training set. For the purposes of the PCA, we evaluate the model at 1000 phases $0 \leq \Delta t \leq 300$ days. Importantly, we only include SNe from the training set when calculating the principal components because our goal is to produce a self-contained classification pipeline that can be applied to any new light curves. The light curves in the test set also tend to be sampled worse than those in the training set, likely because brighter targets were prioritized for spectroscopic follow-up. We then project the light curves in our test set onto the same principal components.

We use the first six PCA coefficients (Figure~\ref{fig:pca}, top), which together explain $>$99.9\% of the sample variance in each filter, plus the peak absolute magnitude (from the model light curve), as features for each single-filter light curve. These features do not have a simple physical interpretation, although coefficients on the first principal component (a generic rising and declining light curve) are strongly correlated with peak magnitude (see Appendix~\ref{sec:importance}). Light curves with a smooth exponential rise and decline (Figure~\ref{fig:pca}, center) are reconstructed nearly perfectly from the principal components, whereas light curves with a plateau (Figure~\ref{fig:pca}, bottom) are not reconstructed as well. Regardless, we find that the PCA coefficients are useful features for classification; a perfect reconstruction of the model light curve is not necessary.

\begin{figure*}
    \centering
    \includegraphics[width=0.75\textwidth]{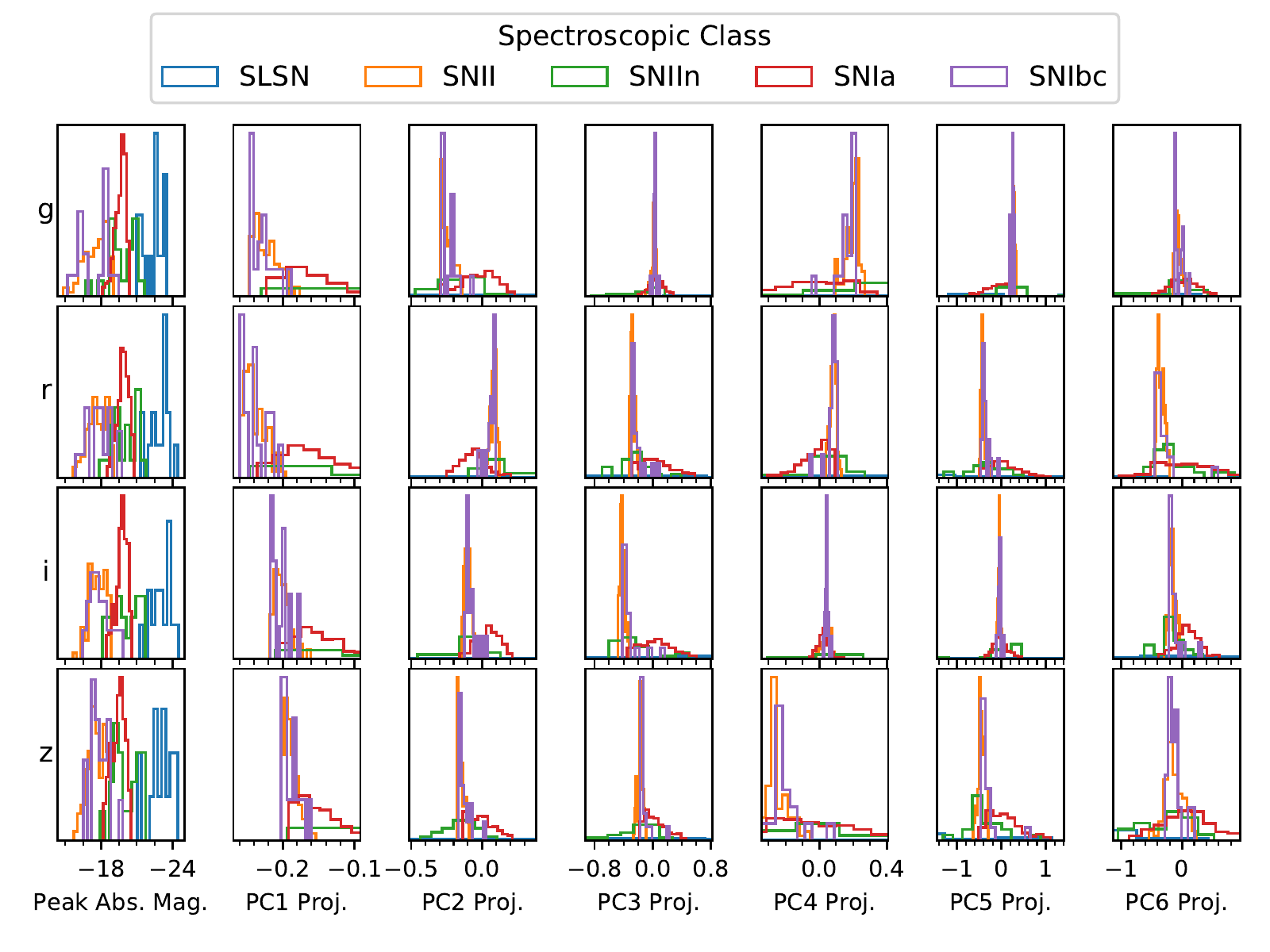}
    \includegraphics[width=0.75\textwidth]{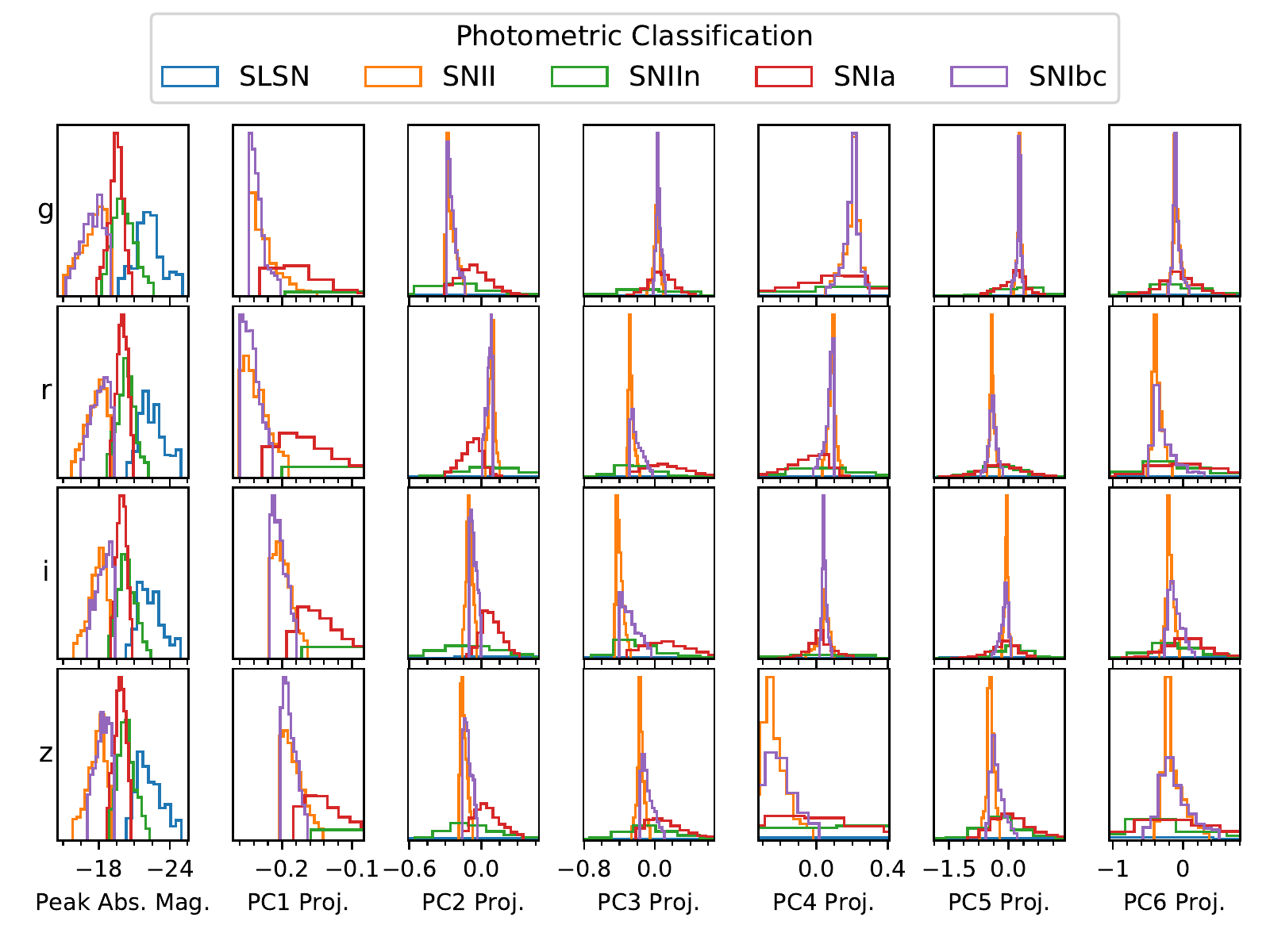}
    \caption{Histograms of the features (before rescaling) for each SN in the training (top) and test (bottom) sets, grouped by spectroscopic (top) and photometric (bottom) classification. Each row shows one filter. SN~IIn and especially SLSN features extend outside the range of the axes. The classes do separate in this feature space. However, the significant overlap between SNe~II and SNe~Ibc, and to a lesser extent between SNe~Ia and SNe~Ibc, makes it difficult to identify members of the less numerous class (SNe~Ibc).}
    \label{fig:features}
\end{figure*}

In total, each multiband SN light curve has 28 features (Figure~\ref{fig:features}). Before classification, we rescale each feature to have zero mean and unit variance in the training set and apply the same scaling to the test set.

\subsection{Classification}
The differences in the observed rates of our five SN classes mean that our training set is unbalanced: \specIa{} Type~Ia, \specII{} Type~II, \specIIn{} Type~IIn, \specIbc{} Type~Ibc, and \specSLSN{} SLSNe. For our classifier to perform effectively on the minority classes, and to increase our sample size in general, we augment each class with additional feature sets by oversampling our training set. \citetalias{villar_supernova_2019} explored two oversampling methods: the synthetic minority oversampling technique (SMOTE; \citealt{chawla_smote_2002}) and multivariate-Gaussian (MVG) oversampling. We implement both options in our code but use the latter for our final classifications, oversampling all classes to have 1000 members. In agreement with \citetalias{villar_supernova_2019}, we find that MVG oversampling gives better results because it allows for features outside the original distribution. We use the implementation of SMOTE in the imbalanced-learn package \citep{lemaitre_imbalanced-learn_2017} and implement our own imbalanced-learn-compatible MVG oversampler based on the \texttt{multivariate\_normal} function in NumPy \citep{oliphant_guide_2006}.

\citetalias{villar_supernova_2019} tested three supervised machine-learning algorithms for classification: a random forest, a support vector machine, and a multilayer perceptron (a type of neural network). Our code includes implementations of all three algorithms from the scikit-learn package \citep{pedregosa_scikit-learn_2011}, but our final classifications use the random forest option with 100 decision trees, an entropy split criterion, and a maximum of five features, as this set of hyperparameters performed best for \citetalias{villar_supernova_2019} (see also Appendix~\ref{sec:hyperparam}). After training the random forest on the oversampled training set, we apply it to the 10 sets of features (from the 10 random posterior draws) for each transient in the test set to get 10 sets of classification probabilities. We then average the 10 sets of classification probabilities for each transient and adopt the classification with the highest probability. 

\begin{figure*}
    \centering
    \includegraphics[height=0.95\textheight]{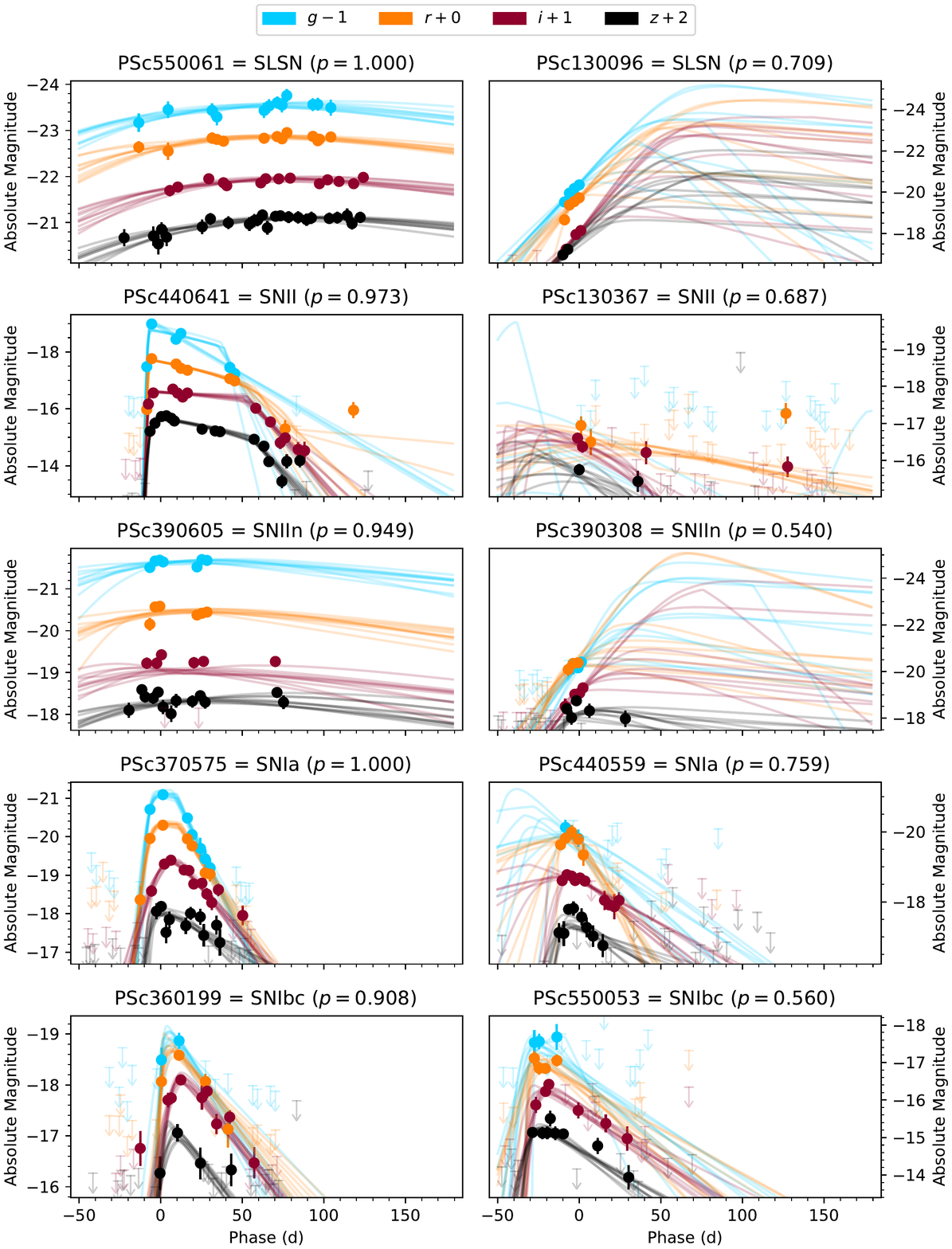}
    \caption{Sample observed (circles) and model (lines) light curves of photometrically classified SNe. Downward-pointing arrows indicate $3\sigma$ upper limits from nondetections. The left column shows the highest confidence classification in each class, and the right column shows the median-confidence classification. Notice that in three of the lower confidence classifications, the light curve is poorly sampled on either the rise or the decline because it was discovered close to the boundaries of the observing season.}
    \label{fig:lcs}
\end{figure*}

\begin{deluxetable*}{cccccCcccccc}
\tablecaption{Classification Results\label{tab:results}}
\tablehead{\colhead{Transient} & \colhead{Host-galaxy} & \colhead{Milky Way} & \colhead{Photometric} & \colhead{Classification} & \colhead{Reduced} & \colhead{Maximum} & \multicolumn{5}{c}{Classification Probabilities} \\[-13pt]
\colhead{} & \colhead{} & \colhead{} & \colhead{} & \colhead{} & \colhead{} & \colhead{} & \multicolumn{5}{c}{------------------------------------------------------} \\[-12pt]
\colhead{Name} & \colhead{Redshift} & \colhead{$E(B-V)$} & \colhead{Classification} & \colhead{Confidence} & \colhead{$\chi^2$} & \colhead{$\hat{R}$} & \colhead{SLSN} & \colhead{SN~II} & \colhead{SN~IIn} & \colhead{SN~Ia} & \colhead{SN~Ibc}}
\startdata
PSc000012 & 0.6226 & 0.0181 & SN~Ia & 0.858 & 1.1 & 1.0 & 0.001 & 0.023 & 0.089 & 0.858 & 0.029 \\
PSc000013 & 0.3720 & 0.0216 & SN~Ibc & 0.843 & 1.2 & 1.0 & 0.000 & 0.002 & 0.024 & 0.131 & 0.843 \\
PSc000015 & 0.2090 & 0.0286 & SN~II & 0.574 & 17.5 & 1.0 & 0.000 & 0.574 & 0.003 & 0.077 & 0.346 \\
PSc000017 & 0.2570 & 0.0213 & SN~II & 0.757 & 2.8 & 1.0 & 0.000 & 0.757 & 0.023 & 0.040 & 0.180 \\
PSc000022 & 0.2530 & 0.0244 & SN~II & 0.665 & 2.5 & 1.1 & 0.000 & 0.665 & 0.015 & 0.062 & 0.258 \\
PSc000031 & 0.2190 & 0.0250 & SN~Ia & 0.592 & 2.9 & 1.0 & 0.002 & 0.024 & 0.375 & 0.592 & 0.007 \\
PSc000032 & 0.1650 & 0.0298 & SN~II & 0.566 & 3.3 & 1.0 & 0.000 & 0.566 & 0.002 & 0.079 & 0.353 \\
PSc000036 & 2.0260 & 0.0196 & SLSN & 0.891 & 1.2 & 1.0 & 0.891 & 0.000 & 0.009 & 0.100 & 0.000 \\
PSc000051 & 0.1940 & 0.0123 & SN~II & 0.838 & -3.2 & 1.1 & 0.000 & 0.838 & 0.010 & 0.042 & 0.110 \\
PSc000059 & 0.7800 & 0.0245 & SN~Ia & 0.384 & 35.5 & 1.0 & 0.376 & 0.006 & 0.224 & 0.384 & 0.010 \\
PSc000060 & 0.1470 & 0.0270 & SN~II & 0.560 & 1.8 & 1.0 & 0.000 & 0.560 & 0.000 & 0.144 & 0.296 \\
PSc000068 & 0.1950 & 0.0256 & SN~II & 0.895 & 2.3 & 1.0 & 0.000 & 0.895 & 0.038 & 0.026 & 0.041 \\
PSc000069 & 0.3360 & 0.0274 & SN~Ia & 0.997 & 1.2 & 1.1 & 0.000 & 0.000 & 0.001 & 0.997 & 0.002 \\
PSc000070 & 0.2030 & 0.0260 & SN~Ibc & 0.480 & 1.1 & 1.0 & 0.000 & 0.456 & 0.001 & 0.063 & 0.480 \\
PSc000075 & 0.0820 & 0.0263 & SN~Ibc & 0.483 & 4.3 & 1.0 & 0.000 & 0.455 & 0.000 & 0.062 & 0.483 \\
PSc000080 & 0.4510 & 0.0283 & SN~Ia & 0.986 & 1.4 & 1.0 & 0.000 & 0.005 & 0.002 & 0.986 & 0.007 \\
PSc000095 & 0.3300 & 0.0289 & SN~Ia & 0.983 & 1.1 & 1.0 & 0.000 & 0.000 & 0.003 & 0.983 & 0.014 \\
PSc000102 & 0.2390 & 0.0143 & SN~Ia & 0.381 & 801.5 & 1.7 & 0.016 & 0.144 & 0.357 & 0.381 & 0.102 \\
PSc000150 & 0.2060 & 0.0092 & SN~Ia & 0.985 & 2.4 & 1.8 & 0.000 & 0.011 & 0.002 & 0.985 & 0.002
\enddata
\tablecomments{This table is available in its entirety in machine-readable form.}
\end{deluxetable*}

\begin{deluxetable*}{ccccccCcccccc}
\tablecaption{Cross-validation Results\label{tab:validation}}
\tablehead{\colhead{Transient} & \colhead{Transient} & \colhead{Milky Way} & \colhead{Spectroscopic} & \colhead{Photometric} & \colhead{Classification} & \colhead{Reduced} & \colhead{Maximum} & \multicolumn{5}{c}{Classification Probabilities} \\[-13pt]
\colhead{} & \colhead{} & \colhead{} & \colhead{} & \colhead{} & \colhead{} & \colhead{} & \colhead{} & \multicolumn{5}{c}{------------------------------------------------------} \\[-12pt]
\colhead{Name} & \colhead{Redshift} & \colhead{$E(B-V)$} & \colhead{Classification} & \colhead{Classification} & \colhead{Confidence} & \colhead{$\chi^2$} & \colhead{$\hat{R}$} & \colhead{SLSN} & \colhead{SN~II} & \colhead{SN~IIn} & \colhead{SN~Ia} & \colhead{SN~Ibc}}
\startdata
PS0909006 & 0.2840 & 0.0426 & SN~Ia & SN~Ia & 0.841 & \infty & 1.7 & 0.004 & 0.017 & 0.092 & 0.841 & 0.046 \\
PS0909010 & 0.2700 & 0.0256 & SN~Ia & SN~Ia & 0.964 & 2.1 & 1.0 & 0.003 & 0.000 & 0.033 & 0.964 & 0.000 \\
PS0910016 & 0.2300 & 0.0219 & SN~Ia & SN~Ia & 0.944 & 2.0 & 1.0 & 0.000 & 0.002 & 0.004 & 0.944 & 0.050 \\
PS0910017 & 0.3200 & 0.0221 & SN~Ia & SN~Ia & 0.984 & 2.1 & 1.0 & 0.000 & 0.002 & 0.008 & 0.984 & 0.006 \\
PS0910018 & 0.2650 & 0.0242 & SN~Ia & SN~Ia & 0.910 & 3.2 & 1.0 & 0.000 & 0.006 & 0.016 & 0.910 & 0.068 \\
PS0910020 & 0.2420 & 0.0130 & SN~Ia & SN~Ia & 0.799 & 1.4 & 1.0 & 0.000 & 0.052 & 0.054 & 0.799 & 0.095 \\
PS0910021 & 0.2560 & 0.0081 & SN~Ia & SN~Ia & 0.845 & 1.3 & 1.0 & 0.000 & 0.013 & 0.142 & 0.845 & 0.000 \\
PSc000001 & 0.0710 & 0.0090 & SN~II & SN~II & 0.734 & -12.9 & 1.7 & 0.000 & 0.734 & 0.030 & 0.095 & 0.141 \\
PSc000006 & 0.2308 & 0.0083 & SN~Ia & SN~Ia & 0.469 & 16.9 & 1.0 & 0.000 & 0.061 & 0.018 & 0.469 & 0.452 \\
PSc000010 & 0.2447 & 0.0224 & SN~Ia & SN~Ia & 0.942 & 2.6 & 1.0 & 0.000 & 0.008 & 0.001 & 0.942 & 0.049 \\
PSc000011 & 0.3800 & 0.0177 & SN~Ia & SN~Ibc & 0.683 & 1.0 & 1.0 & 0.000 & 0.000 & 0.007 & 0.310 & 0.683 \\
PSc000014 & 0.1369 & 0.0261 & SN~Ia & SN~Ia & 0.883 & 8.6 & 1.1 & 0.000 & 0.014 & 0.077 & 0.883 & 0.026 \\
PSc000034 & 0.2500 & 0.0278 & SN~Ia & SN~Ia & 0.706 & 1.3 & 1.0 & 0.011 & 0.021 & 0.244 & 0.706 & 0.018 \\
PSc000038 & 0.1500 & 0.0220 & SN~Ia & SN~Ia & 0.996 & 1.6 & 1.0 & 0.000 & 0.001 & 0.001 & 0.996 & 0.002 \\
PSc000076 & 0.2600 & 0.0245 & SN~II & SN~II & 0.757 & 1.1 & 1.0 & 0.000 & 0.757 & 0.004 & 0.049 & 0.190 \\
PSc000091 & 0.1520 & 0.0269 & SN~Ia & SN~Ia & 0.599 & 3.1 & 2.9 & 0.000 & 0.005 & 0.003 & 0.599 & 0.393 \\
PSc000098 & 0.0570 & 0.0127 & SN~II & SN~II & 0.673 & 4.8 & 3.0 & 0.000 & 0.673 & 0.009 & 0.049 & 0.269 \\
PSc000133 & 0.2440 & 0.0082 & SN~II & SN~Ia & 0.373 & 3.4 & 1.0 & 0.000 & 0.285 & 0.183 & 0.373 & 0.159 \\
PSc000137 & 0.1183 & 0.0080 & SN~Ia & SN~Ia & 0.780 & 2.8 & 1.0 & 0.000 & 0.005 & 0.003 & 0.780 & 0.212
\enddata
\tablecomments{This table is available in its entirety in machine-readable form.}
\end{deluxetable*}

\section{Results and Validation}\label{sec:validate}
Applying our classification pipeline to the test set described in Section~\ref{sec:data} yields \photIa{} photometrically classified SNe~Ia, \photII{} SNe~II, \photIbc{} SNe~Ibc, \photIIn{} SNe~IIn, and \photSLSN{} SLSNe.\footnote{To exactly reproduce our results, a seed of 0 must be used in the pseudorandom number generators during feature extraction, oversampling, and classification.} Table~\ref{tab:results} gives the full list of classifications, and Figure~\ref{fig:lcs} shows a sample of photometrically classified light curves. These are among the largest samples of each of these classes of SNe in the literature from a single survey. In the remainder of this section, we assess the performance of our algorithm in general and discuss how to use these classifications in practice.

\subsection{Cross-validation}\label{sec:crossvalidate}
\begin{figure}
    \centering
    \includegraphics[width=0.91\columnwidth]{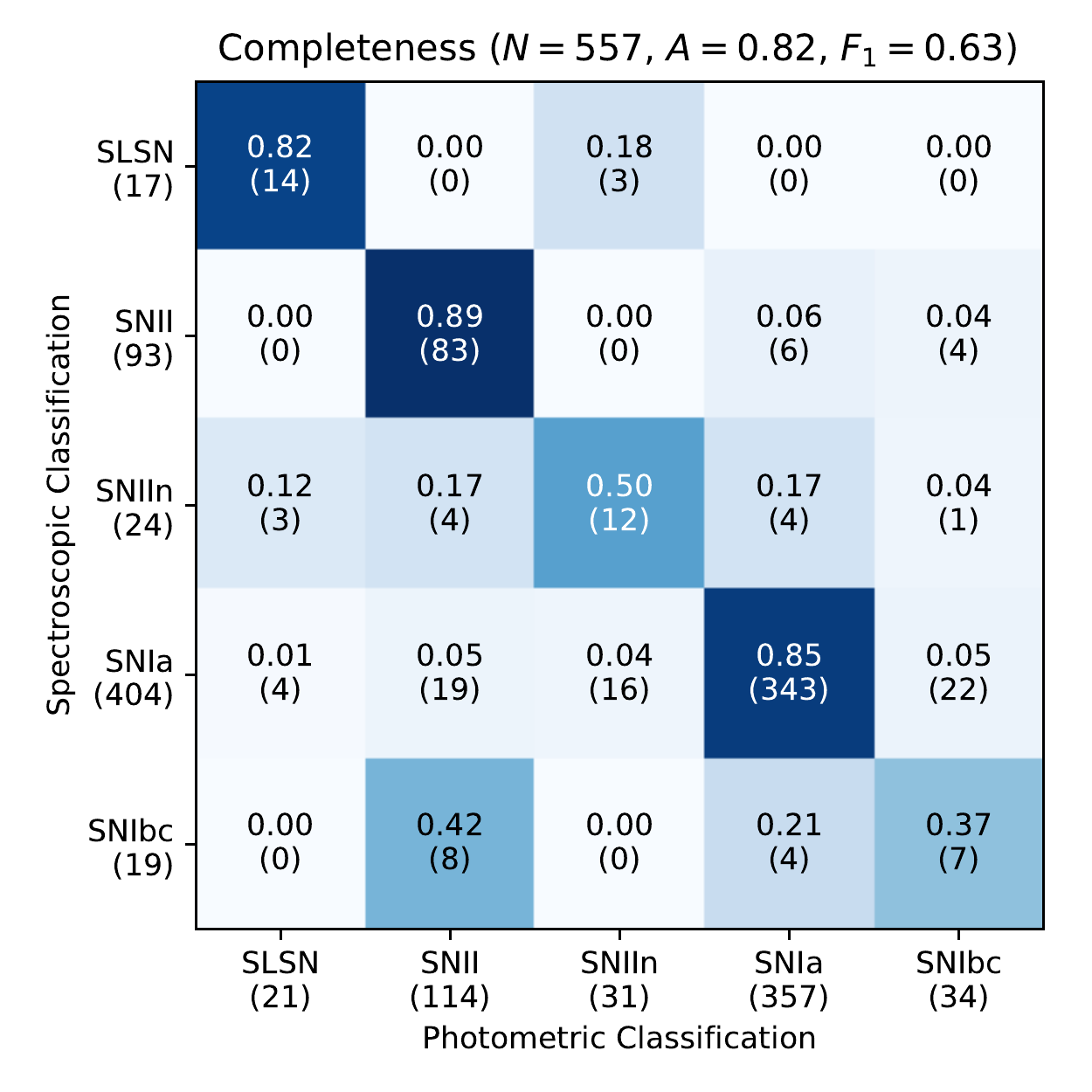}
    \includegraphics[width=0.91\columnwidth]{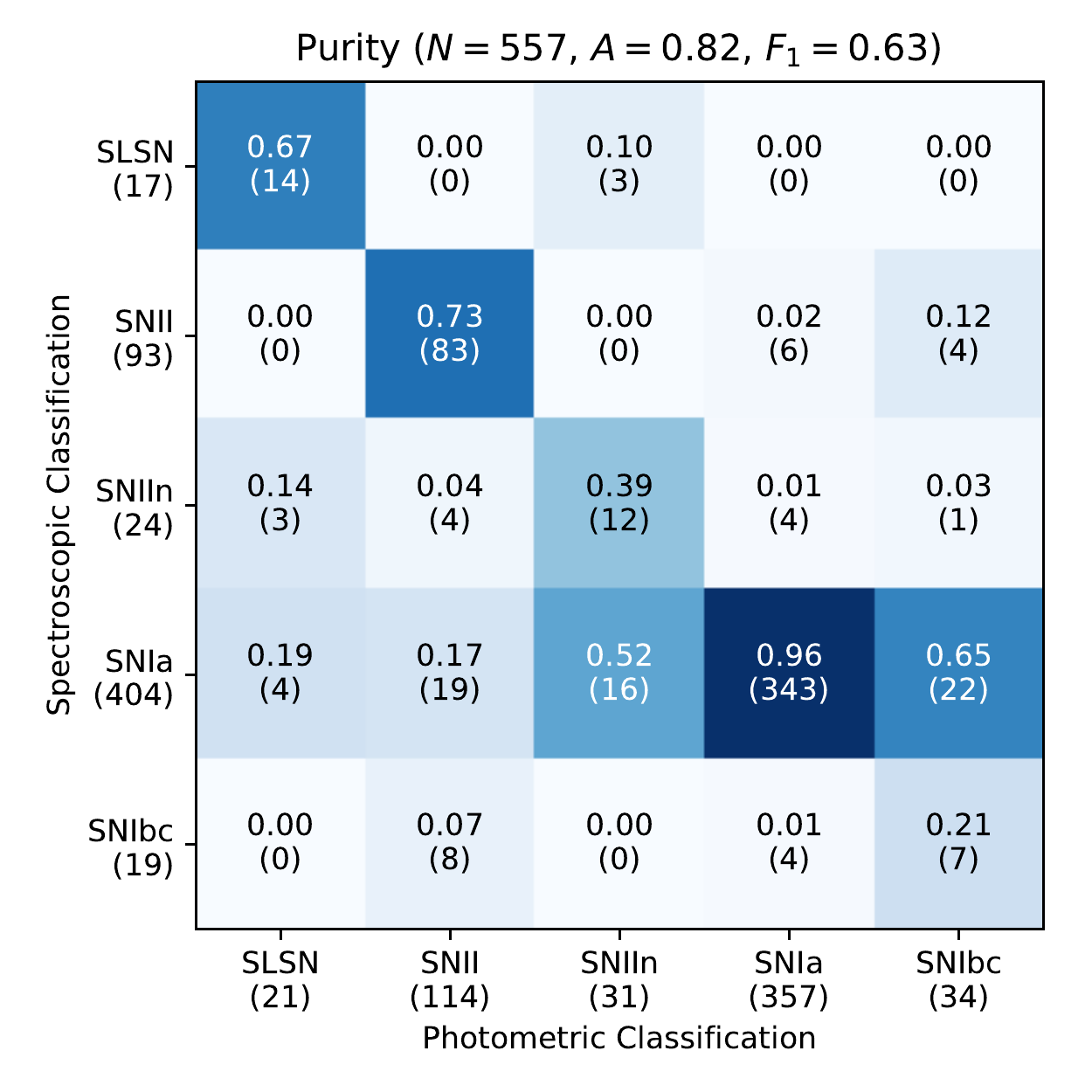}
    \includegraphics[width=0.45\columnwidth]{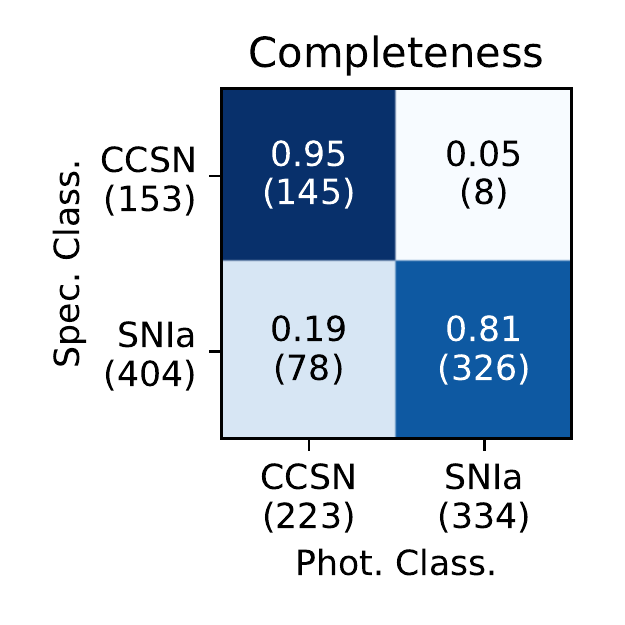}
    \includegraphics[width=0.45\columnwidth]{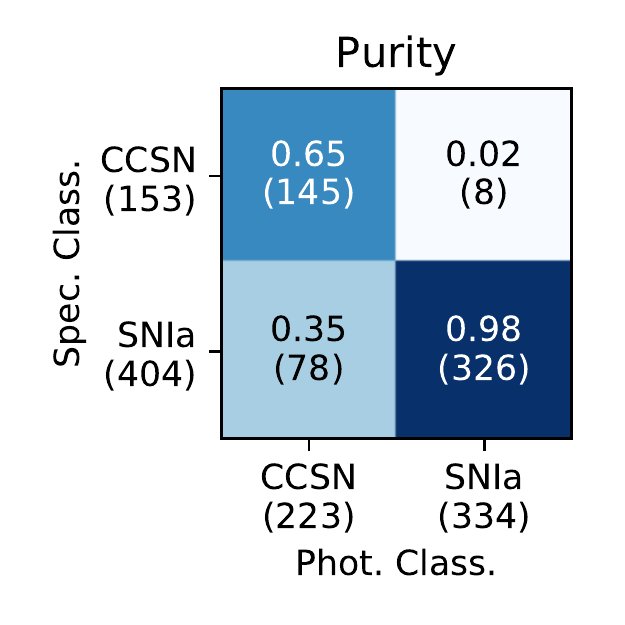}
    \caption{Confusion matrices of our validation results. Each cell lists and is colored by the fraction of each class, with the raw number in parentheses. The top matrix aggregates by true label, so its diagonal shows completeness. The middle matrix aggregates by predicted label, so its diagonal shows purity. The bottom matrices are the same results but with all four non-SN~Ia classes grouped together as CCSNe. $N$ is the total size of the training set, $A$ is the accuracy, and the $F_1$ is a class-weighted average of completeness and purity (see Appendix~\ref{sec:glossary} for definitions).}
    \label{fig:confusion}
\end{figure}

We first validate our classifier using leave-one-out cross-validation with the SNe in our training set. For each iteration of the cross-validation, we retrain the classifier on all but one of the SNe in our training set (still using the median parameters) and then use it to classify 10 sets of features derived from the posterior parameter distributions for that SN. We then average the 10 sets of probabilities to determine the cross-validation classifications listed in Table~\ref{tab:validation}. Figure~\ref{fig:confusion} shows the resulting confusion matrices.

In general our code performs well, with an overall classification accuracy of \accuracy{}. (See Appendix~\ref{sec:glossary} for a glossary of terms.) This is dominated by the photometrically homogeneous SNe~Ia, which are \completenessIa{} complete. SLSNe, which are \completenessSLSN{} complete, are also easy to identify because they separate relatively cleanly by absolute magnitude. SNe~IIn (\completenessIIn{} complete) and Ibc (\completenessIbc{} complete) are hardest to identify because their light curves are intrinsically more heterogeneous, and because we have relatively small numbers of each in the training set, so their diversity is not well sampled. SNe~Ibc also overlap significantly with SNe~II in feature space (Figure~\ref{fig:features}).

In addition to being complete, most of our photometric classes are relatively pure: \purityIa{} for SNe~Ia, \purityII{} for SNe~II, and \puritySLSN{} for SLSNe. For SNe~IIn and Ibc, our purity falls to \purityIIn{} and \purityIbc{}, respectively. The latter raises an important point: in a magnitude-limited survey, where SNe~Ia make about 70\% of observed SNe, a small fraction of misclassified SNe~Ia can significantly contaminate rarer classes. For example, in our data set, the \cmIaIbc{} of SNe~Ia misclassified as SNe~Ibc represent \pmIaIbc{} of our photometrically classified SNe~Ibc.

The most common misclassification is labeling \cmIbcII{} of SNe~Ibc as SNe~II. Intriguingly, misclassifications in the opposite direction happen at much lower rates. This shows that the random forest has labeled a larger region of feature space as SN~II (see also Figure~\ref{fig:params}). We suspect this is due to the fact that our SN~II training set includes light curves with very flat plateaux as well as steep exponential declines (previously referred to as SNe~IIP and IIL, respectively).

If we aggregate all classes other than SNe~Ia under the label core-collapse SN (CCSN), this combined class is \completenessCCSN{} complete and \purityCCSN{} pure (Figure~\ref{fig:confusion}, bottom). This shows that most of our misclassifications are between subtypes of CCSNe, rather than between CCSNe and SNe~Ia. In the case of the binary classification, our sample of SNe~Ia is actually purer (\purityIaBinary{} vs.\ 94\%), but less complete (\completenessIaBinary{} vs.\ 91\%), than the final photometric sample of SNe~Ia from the SDSS-II SN Survey \citep{sako_photometric_2011}.

\subsection{Class Fractions}
If we ignore any biases in selecting targets for spectroscopic follow-up,\footnote{Our spectroscopic follow-up program serviced multiple science goals, so we consider the spectroscopic class fractions to be roughly representative of a magnitude-limited survey.} the true class fractions in the photometrically classified sample should approximately match the class fractions in the spectroscopically classified sample. However, our algorithm has different misclassification rates for each class of SNe, which we have measured using cross-validation. We can test the validity of these measured misclassification rates by using them to ``correct'' the class fractions in the photometrically classified sample and checking if the corrected fractions match the fractions in the spectroscopically classified sample. 

\begin{figure}
    \centering
    \includegraphics[width=\columnwidth]{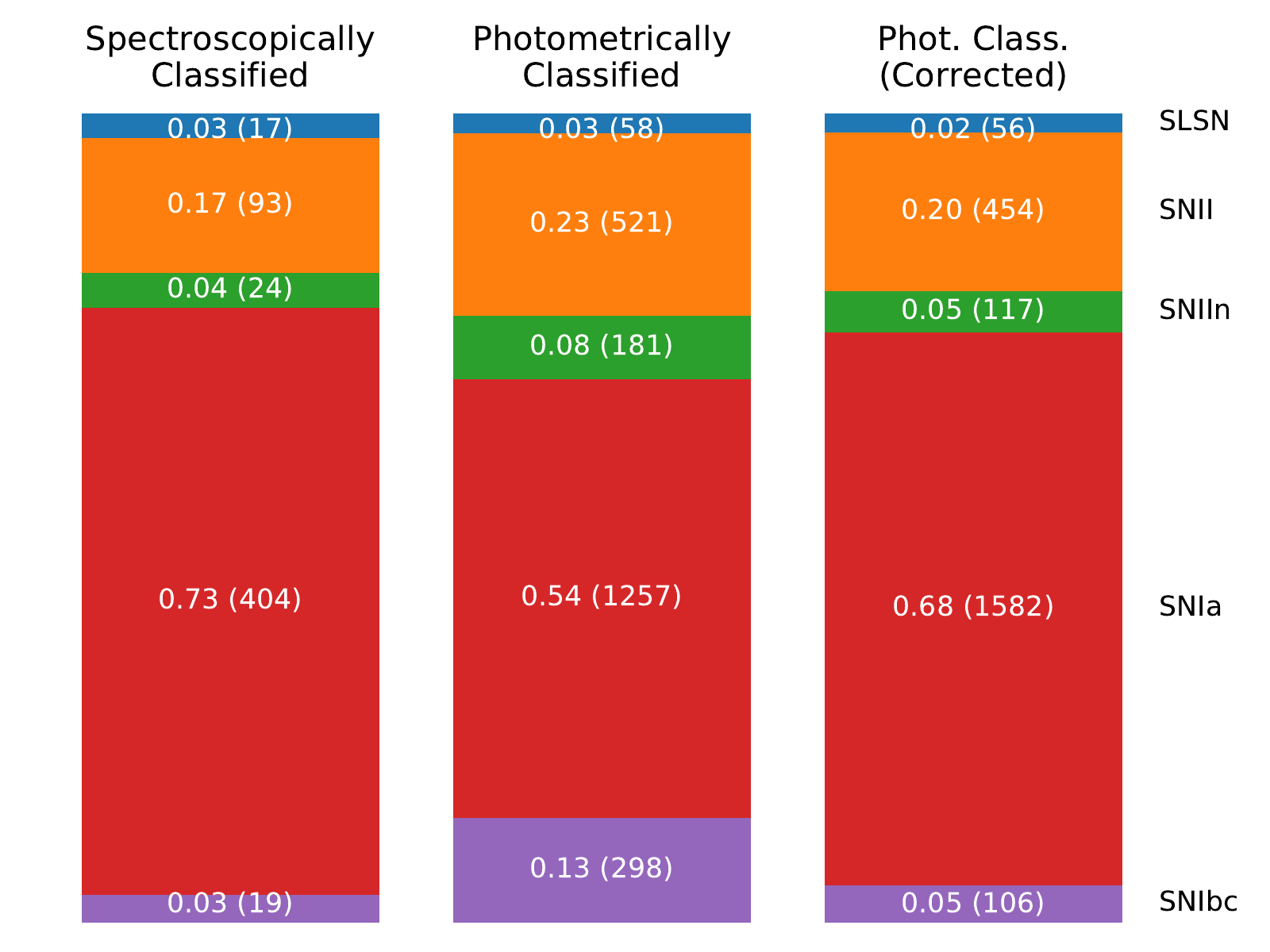}
    \caption{Fractions of our spectroscopic (left) and photometric (center) samples belonging to each class. The difference in class breakdown is likely due to small fractions of SNe~Ia being misclassified as SNe~II and Ibc. The ``corrected'' fractions (right) account for our expected misclassification rates (i.e., Figure~\ref{fig:confusion}). Under the assumption that we performed random spectroscopic follow-up, similarity between the composition of our spectroscopic and corrected photometric samples would suggest that we understand the performance of our classifier.}\label{fig:bar}
\end{figure}

For example, we observe that our photometrically classified sample contains a larger fraction of SNe~II and Ibc and a smaller fraction of SNe~Ia compared to our spectroscopically classified sample (Figure~\ref{fig:bar}). From our confusion matrix (Figure~\ref{fig:confusion}, center), we can see that this is due to small fractions of SNe~Ia contaminating the photometric SN~II and Ibc samples. If we correct these fractions for the measured misclassification rates in our training set, we obtain a class breakdown similar to our training set (Figure~\ref{fig:bar}, center). This suggests that the numbers in our confusion matrix are a good representation of the performance of our classifier.

\subsection{Confidence Thresholds}
\begin{figure*}
    \centering
    \includegraphics[width=\columnwidth]{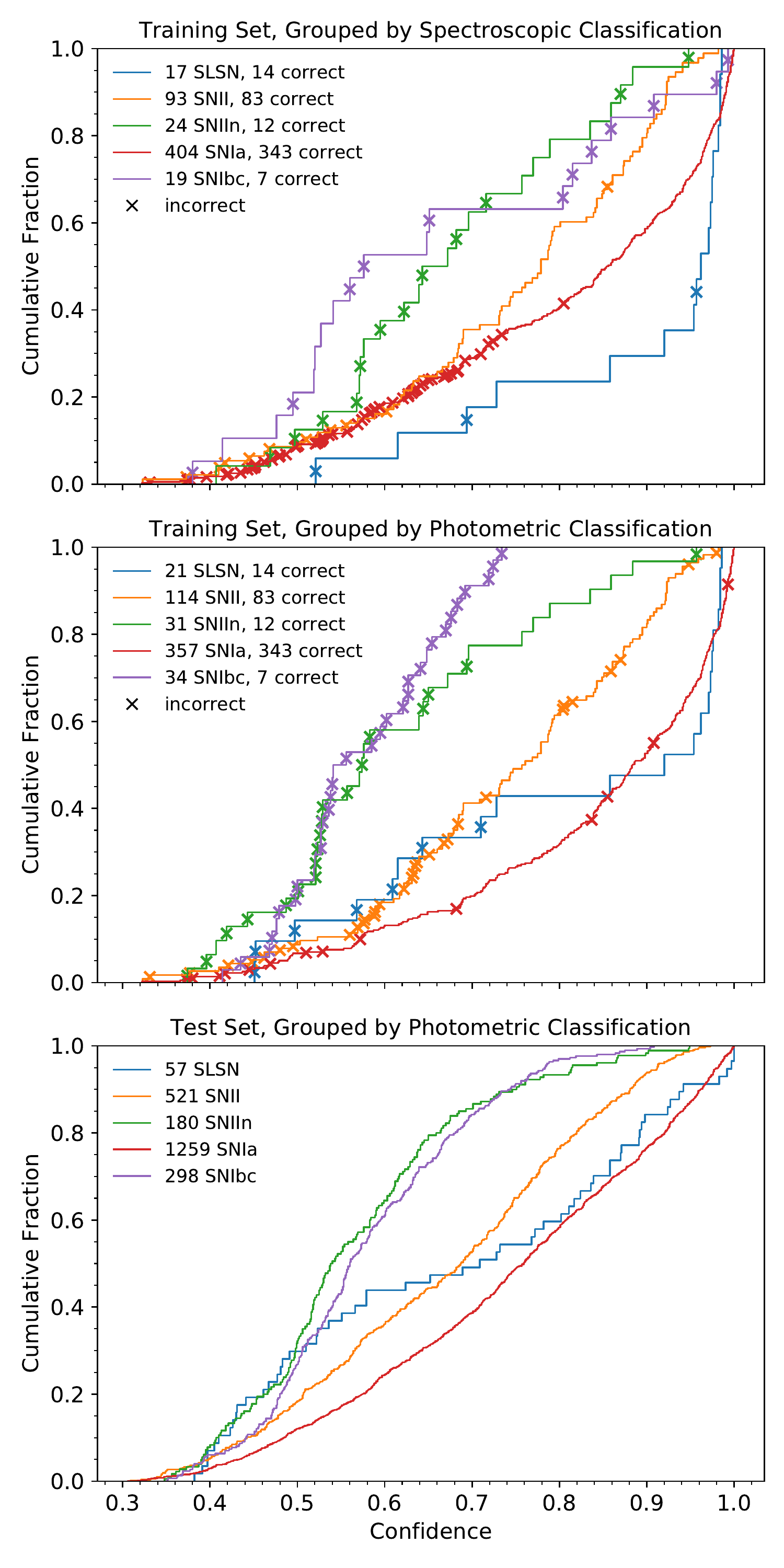}
    \includegraphics[width=\columnwidth]{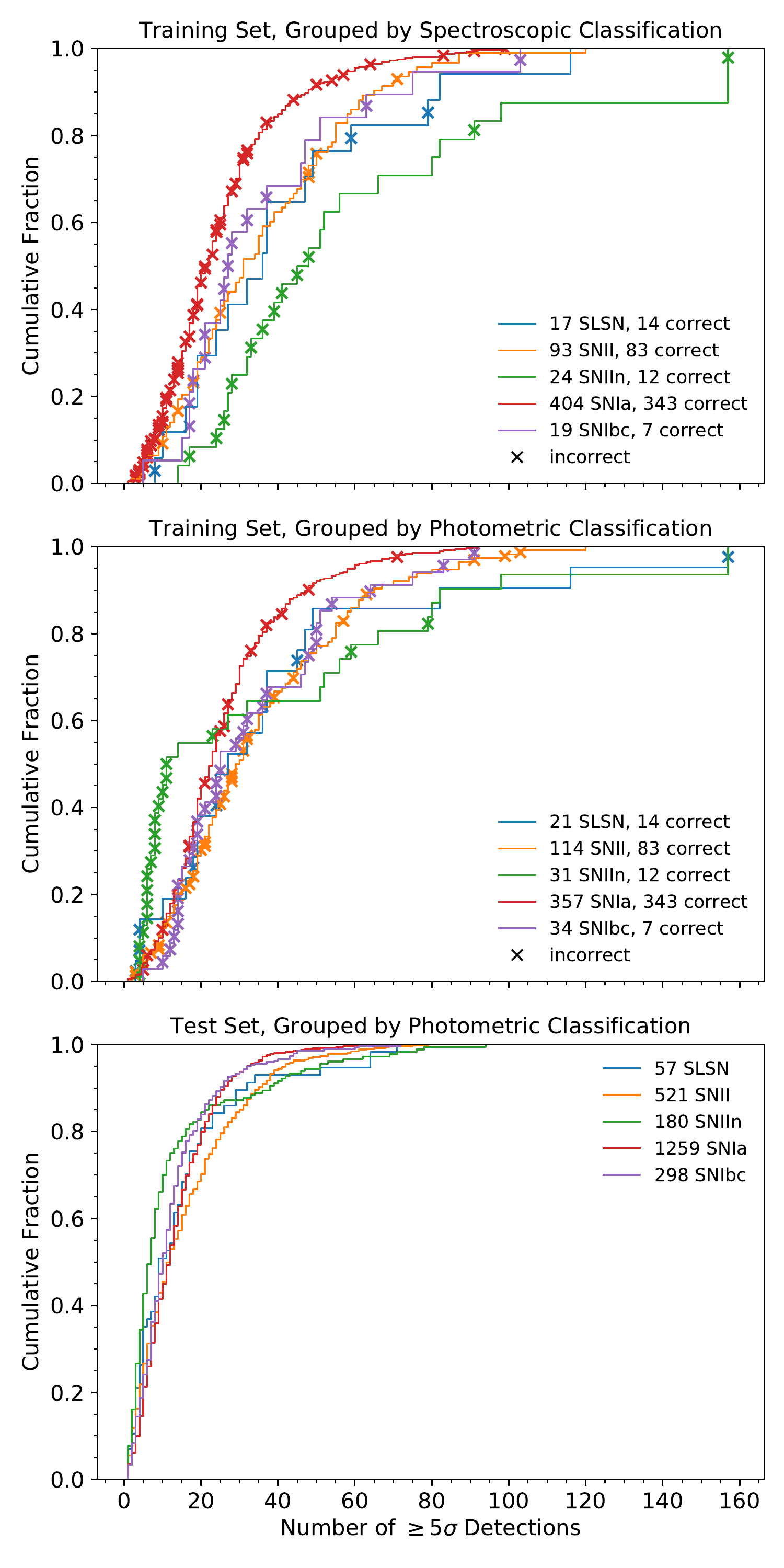}
    \caption{Cumulative histograms of the classification confidence (left column) and the number of $\geq\!5\sigma$ detections (right column) for the training set (top and center rows) and the test set (bottom row), grouped by spectroscopic classification (top row) or photometric classification (center and bottom rows). Transients whose photometric and spectroscopic classifications do not match are marked by an $\times$. SLSNe and SNe~Ia are typically classified with the highest confidence. With the exception of SNe~Ibc, most false-positive classifications have low confidence and/or few detections.}
    \label{fig:probabilities}
\end{figure*}

We also assess how confident our classifier is in its predictions by examining the classification confidence (the highest classification probability) for each SN in the training set. Figure~\ref{fig:probabilities} shows cumulative histograms of the classification confidences for each spectroscopic class (top left) and photometric class (center left). As expected, we find that on average the classifier is most confident in predicting SLSNe and SNe~Ia, and least confident in SNe~Ibc and IIn. In fact, 9 of the 12 highest confidence ($p > 0.8$) misclassifications are for spectroscopically classified SNe~Ibc and IIn. We cannot assess the correctness of the classifications in the test set, but we observe that the distributions of the classification confidence in the test set (Figure~\ref{fig:probabilities}, bottom left) are similar to those in the training set. This suggests that our claims about misclassification rates may generalize to the test set.

\begin{figure}
    \centering
    \includegraphics[width=\columnwidth]{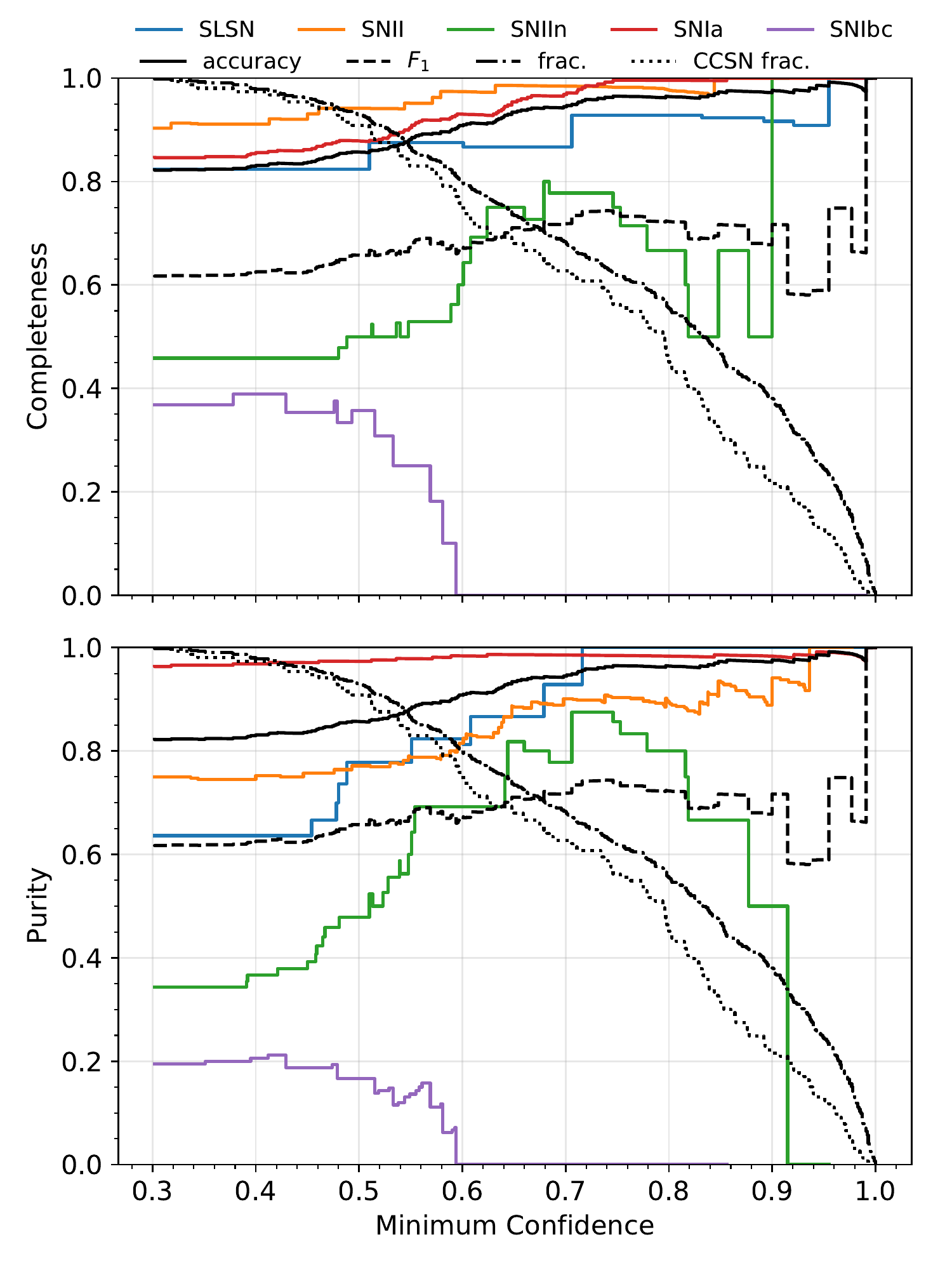}
    \includegraphics[width=\columnwidth]{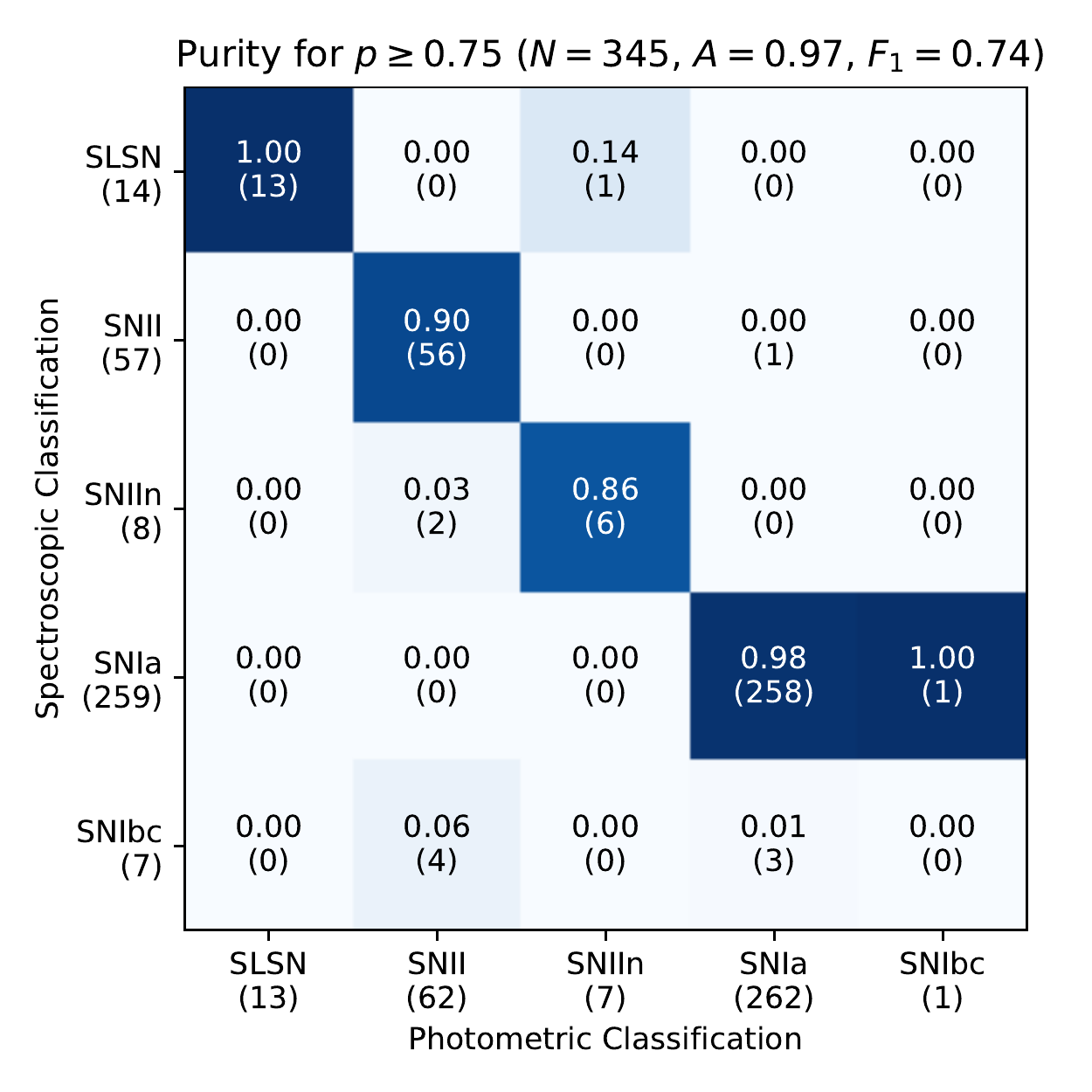}
    \caption{Completeness (top) and purity (center) for each class as a function of the minimum acceptable confidence. The total accuracy (solid), $F_1$ score (dashed), fraction of events remaining (dotted-dashed), and fraction of CCSNe remaining (dotted) are shown in black on both panels for reference. By only considering photometrically classified events above a certain threshold (e.g., \confidenceThresh{}, bottom), we can increase the purity of most of our samples, at the expense of decreasing the absolute number of events (in this case by \exclusionAtThresh{}).}
    \label{fig:threshold}
\end{figure}

We can increase the completeness and/or purity of our photometrically classified samples by considering only transients classified with confidence above a certain threshold, at the expense of decreasing their absolute numbers. Figure~\ref{fig:threshold} (top and center) shows how our performance metrics vary as a function of the confidence threshold chosen. There is no clear optimum for all classes, so any threshold is arbitrary. However, a threshold of \confidenceThresh{}, for example, is better than using the full sample for all classes but SNe~Ibc, and only excludes \exclusionAtThresh{} of the training set (\exclusionAtThreshCCSN{} of CCSNe). Figure~\ref{fig:threshold} (bottom) shows the confusion matrix that results from imposing this threshold. With the exception of SNe~Ibc (which are nearly eliminated), all classes are over \completenessAtThresh{} complete.

Likewise, we can improve our photometrically classified samples by requiring a certain number of photometric observations in order to remove poorly sampled light curves. Figure~\ref{fig:probabilities} shows analogous histograms of the number of $\geq\!5\sigma$ detections (in all bands) for each spectroscopic (top right) and photometric (center right) class. We do find that misclassifications are more frequent for poorly sampled light curves, but the number of points required for a correct classification varies significantly between the spectroscopic classes. For example, most SNe~IIn with fewer than 50 observations are misclassified, whereas most SNe~Ia and SNe~II require only 20--30 points. Light curves with fewer than 10 detections ($\sim$2 per filter) are almost always misclassified. This is close to the median number of detections in our test set (Figure~\ref{fig:probabilities}, bottom right). Because of the clear difference in the distributions of the training and test sets, we do not know if a threshold chosen for the training set will have the desired effect on the test set. We also find that a confidence threshold is more effective and removes a smaller fraction of the sample, so we do not adopt a threshold on the number of detections.

\begin{figure*}
    \centering
    \includegraphics[width=0.49\textwidth]{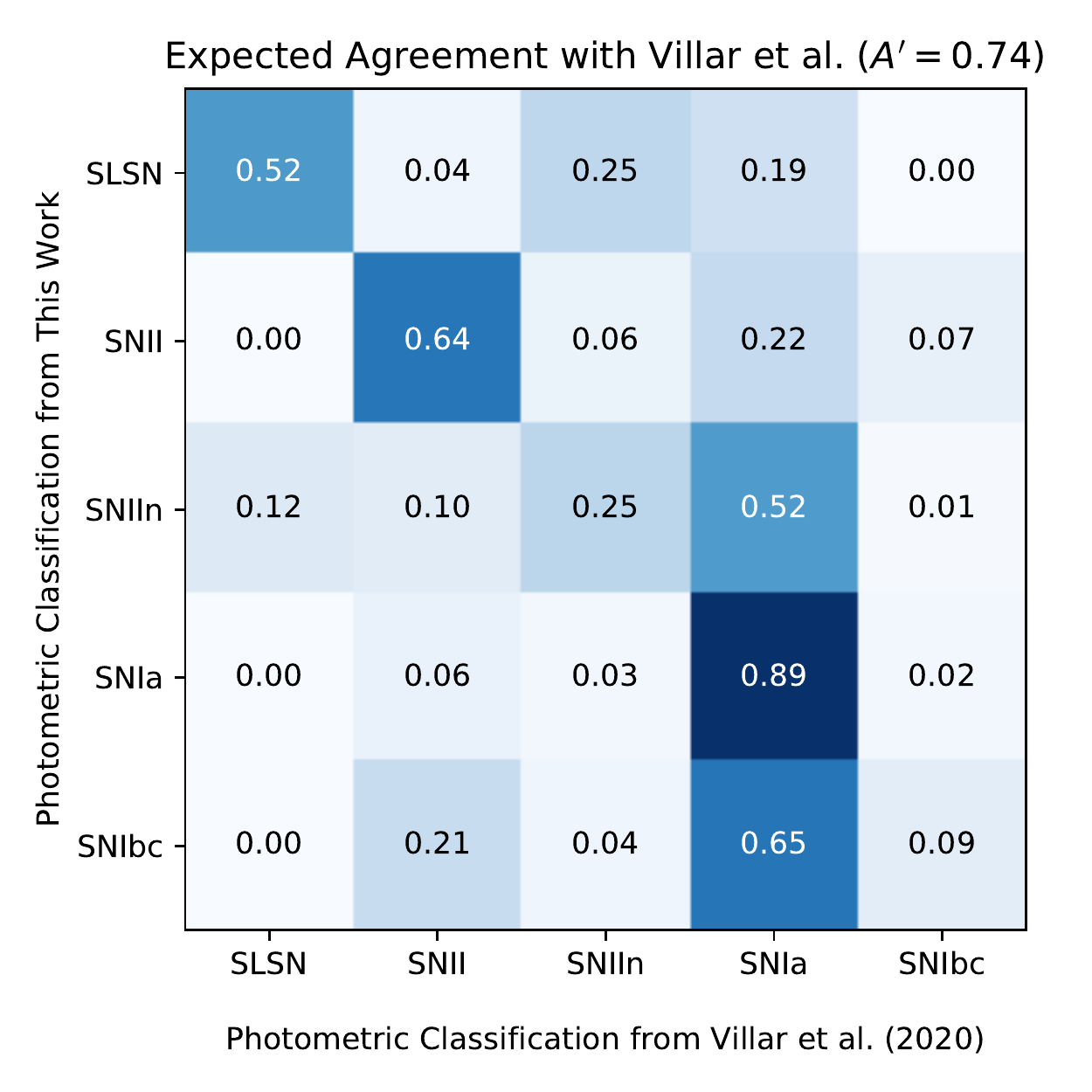}
    \includegraphics[width=0.49\textwidth]{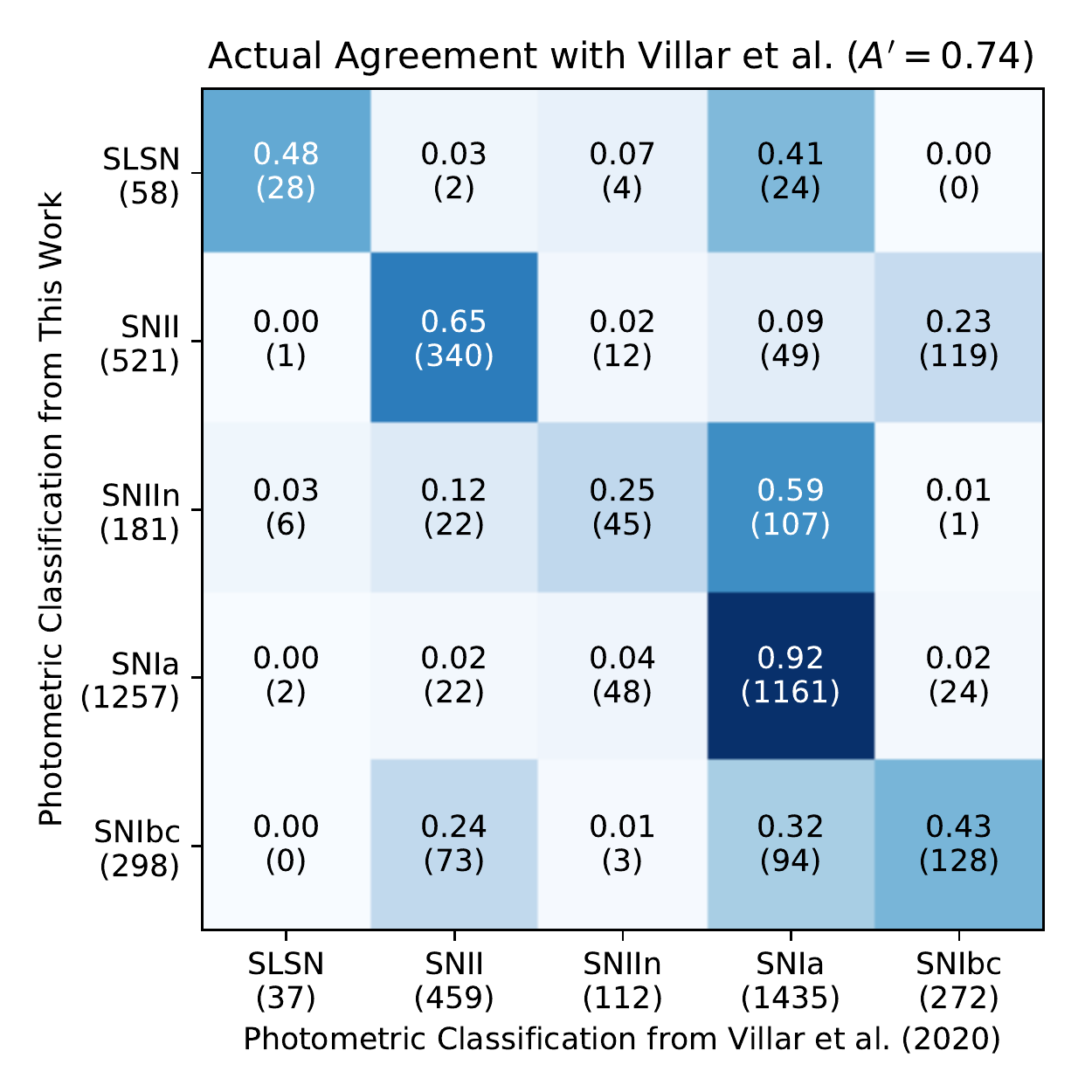}
    \includegraphics[width=0.75\textwidth]{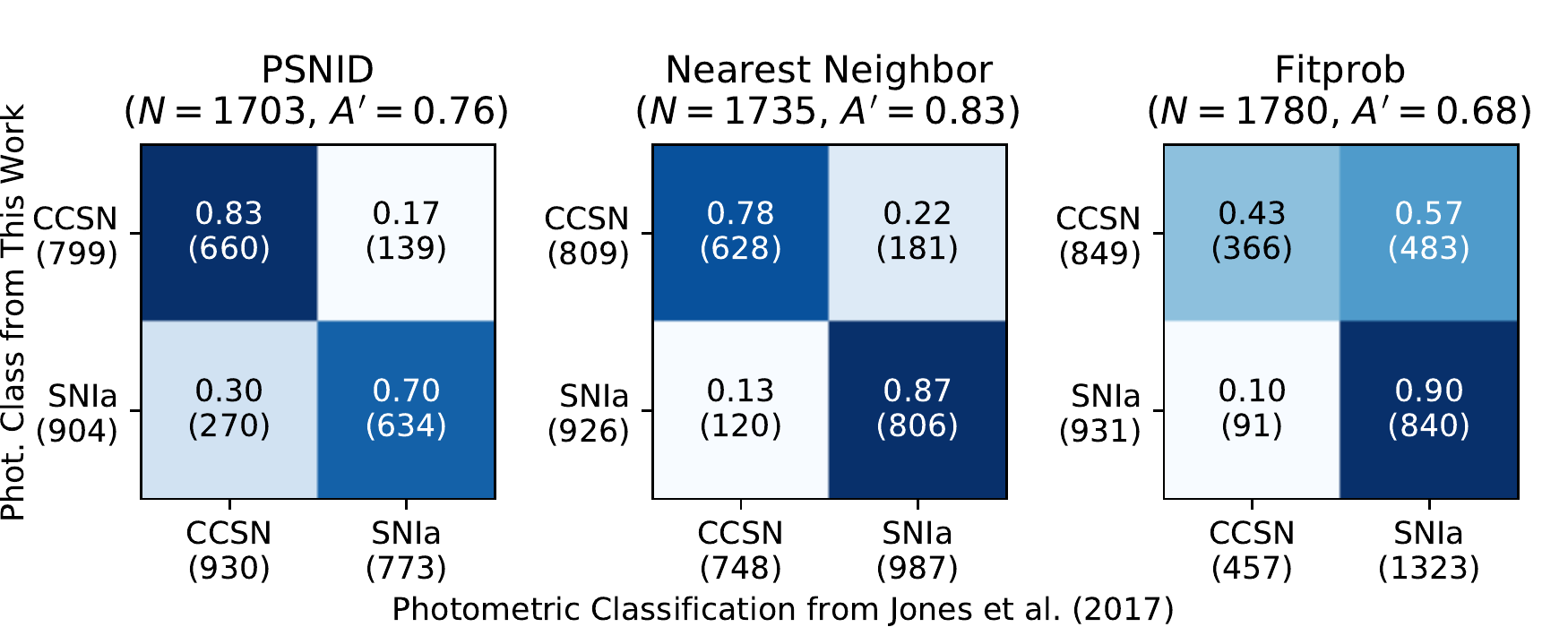}
    \caption{Top left: expected agreement between our classifier and that of \citetalias{villar_superraenn_2020}, given our confusion matrices from validation (see Appendix~\ref{sec:agree}). We expect very low agreement on SNe~IIn and Ibc, which both classifiers independently struggle with. Top right: actual agreement with \citetalias{villar_superraenn_2020}. As expected, we agree on $A' = \agreeVillar{}$ of classifications overall, including a larger than expected fraction of SNe~Ibc (\agreeIbc{}). Bottom: agreement between our classifier and those of \cite{jones_measuring_2017}, where all classes that are not SNe~Ia are aggregated under the CCSN label.}
    \label{fig:agree}
\end{figure*}

\subsection{Comparison to Other Photometric Classifiers}\label{sec:compare}
Although we cannot judge the correctness of individual classifications in our test set, we can see how often our algorithm agrees with other photometric classifiers applied to the same data set. In general, if two classifiers are independent for a given transient, we cannot expect their agreement matrix to be much better than the product of their confusion matrices (see Appendix~\ref{sec:agree} for a full derivation). Each has its own strengths and weaknesses, which can be assessed through validation. 

In particular, we compare to the semisupervised machine-learning classifier of \citetalias{villar_superraenn_2020}, SuperRAENN, which is somewhat less accurate than our classifier but has advantages in terms of speed and extensibility (see \citetalias{villar_superraenn_2020} for more details). We also compare to three CCSN versus SN~Ia classifiers applied by \cite{jones_measuring_2017}: the Photometric Supernova Identification (PSNID; \citealt{sako_photometric_2011}) code provided in the Supernova Analysis (SNANA; \citealt{kessler_snana_2009}) package, which compares light curves to templates of SNe~II, SNe~Ia, and SNe~Ibc; and the ``Nearest Neighbor'' and ``Fitprob'' classifiers, which compare light curves to the SALT2 SN~Ia template \citep{jha_improved_2007} in parameter space and flux space, respectively. Figure~\ref{fig:agree} shows our ``agreement matrices'' with these classifiers. The numbers in the title of each panel indicate the number of transients we have in common ($N$) and the fraction of classifications we agree on ($A'$), and the diagonals show the fraction of our classifications that they agree with for each class.

As expected, \citetalias{villar_superraenn_2020} agree with \agreeVillar{} of our classifications overall, including large fractions of classes that we classify most accurately: \agreeIa{} of SNe~Ia, \agreeII{} of SNe~II, and \agreeSLSN{} of SLSNe. (Our agreement on cross-validation predictions on the training set, \agreeVillarTrain{} overall, follows a similar pattern.) Figure~\ref{fig:agree} (top) reveals three important trends. First, \citetalias{villar_superraenn_2020} agree with \agreeIbc{} of our SNe~Ibc, almost five times more than expected from from our cross-validation results. This implies that both of our classifiers tend to misclassify the same types of transients as SNe~Ibc (i.e., they break the assumption of independence in Appendix~\ref{sec:agree}). Second, \citetalias{villar_superraenn_2020} tend to classify many more transients as SNe~Ia than we do. This may be a result of the imbalanced training set, which their method cannot fully account for (see \citetalias{villar_superraenn_2020} for further discussion). Third, they classify \amSLSNIaabs{} of our SLSNe (\amSLSNIa{}) as SNe~Ia. This is surprising because both SLSNe and SNe~Ia are relatively easy to identify photometrically.

A visual inspection of these light curves shows that most are missing either the rise or decline, due to the beginning or end of the observing season for that field. This appears to lead to a failure mode where some of the model light curves peak around SLSN luminosities, even if none of the observed photometry is that bright (see, e.g., Figure~\ref{fig:lcs}, top right). In a way, this is the desired behavior: if the peak is not observed, we want to include the possibility that the transient may peak at a flux brighter than the brightest observed data point. However, this introduces a bias toward higher luminosities, because the model light curves will never peak significantly below the brightest observed point. This is one reason we introduced a log-uniform prior on the amplitude, a modification to the method of \citetalias{villar_supernova_2019}.

\cite{jones_measuring_2017} also agree with \agreeFitprob{} (Fitprob) to \agreeNN{} (Nearest Neighbor) of our classifications, where we have again aggregated all classes other than SNe~Ia under the label CCSN. However, they had a more specific goal than we do: to produce a pure sample of SNe~Ia for the purpose of measuring cosmological parameters. Therefore, we expect their SN~Ia samples to be less complete and their CCSN samples to be less pure than our samples. Our agreement matrix with PSNID, which \cite{jones_measuring_2017} adopt as their preferred classifier, can indeed be interpreted as reflecting their preference for SN~Ia purity over CCSN purity.

When deciding between two conflicting classifications, one should take into account the purity of the samples produced by each classifier, as well as the relative rates of the two classes in a magnitude-limited survey. For example, for the \amIbcIaabs{} transients that we classify as SNe~Ibc and \citetalias{villar_superraenn_2020} classify as SNe~Ia, we prefer the SN~Ia classification for the majority of them, because these are much more common in nature and we know that our SN~Ibc purity is only \purityIbc{}. Of course, this type of analysis also depends on one's science goals (e.g., purity or completeness of a sample).

\section{Discussion}\label{sec:discuss}
\subsection{Future Applications}
Much of the success of our algorithm (unlike, e.g., \citealt{jones_measuring_2017,jones_measuring_2018}) depends on the discriminating power of absolute magnitude, for which we need a redshift. While we still used spectroscopy to determine these redshifts, most of them were determined after the transients had faded by observing their host galaxies with a multifiber spectrograph, which is much less time consuming than classification spectroscopy of one transient at a time. However, this is still not scalable to the sample sizes expected from LSST. The performance of the algorithm has not yet been tested with photometric redshifts. \cite{graham_photometric_2018} suggest that LSST will determine photometric redshifts to $\sim$5\% accuracy for galaxies with $r \lesssim 25$~mag within the first two years of survey operations. This would be only a small contribution to our classification uncertainties.

Furthermore, our classifications rely on the full light curves of these transients, in contrast to other algorithms that aim to classify transients in real time (e.g., \citealt{muthukrishna_rapid_2019,sravan_real-time_2020}). Our code will therefore be most successful at the end of an observing season, when the user has a large training sample of spectroscopically classified transients in hand and wants to make scientific use of the remaining transients with light curves only. However, our Bayesian light-curve modeling allows for fitting only part of the light curve while keeping all possible future behavior within the parameter uncertainties. Future work will explore what fraction of the light curve is required for good results.

Lastly, increasing the size of the training set would likely improve our results significantly, especially for SNe~Ibc. With so few examples to train on, the algorithm is very sensitive to including or excluding even single events, as demonstrated by our low cross-validation scores for that classes (Figure~\ref{fig:confusion}). However this is not a shortcoming of the algorithm, but rather a reflection of the scarcity of large SN~Ibc samples in the literature \citep{bianco_multi-color_2014,taddia_early-time_2015,taddia_carnegie_2018,stritzinger_carnegie_2018}.

Our current classifier returns probabilities determined entirely by the photometric data. In principle, we could multiply these probabilities by the relative rates of the various classes of SNe observed in previous surveys (e.g., \citealt{graur_loss_2017_1,graur_loss_2017_2,fremling_zwicky_2019,holoien_asas-sn_2019}). This would have the effect of biasing borderline cases toward a more common classification; for example, it would decrease the number of transients that we classify as SLSNe and \citetalias{villar_superraenn_2020} classify as SNe~Ia. We choose not to adopt such a prior in this work so that future analyses have the option of adopting the rate measurements of their choice. In practice, users may also want to combine photometric classification with contextual classification (e.g., \citealt{foley_classifying_2013,baldeschi_star_2020,gomez_fleet_2020}; N.~Chou et al.\ 2020, in preparation).

\begin{deluxetable*}{ccccccCcccccc}
\tablecaption{Rare Transients\label{tab:other}}
\tablehead{\colhead{Transient} & \colhead{Transient} & \colhead{Milky Way} & \colhead{Spectroscopic} & \colhead{Photometric} & \colhead{Classification} & \colhead{Reduced} & \colhead{Maximum} & \multicolumn{5}{c}{Classification Probabilities} \\[-13pt]
\colhead{} & \colhead{} & \colhead{} & \colhead{} & \colhead{} & \colhead{} & \colhead{} & \colhead{} & \multicolumn{5}{c}{------------------------------------------------------} \\[-12pt]
\colhead{Name} & \colhead{Redshift} & \colhead{$E(B-V)$} & \colhead{Classification} & \colhead{Classification} & \colhead{Confidence} & \colhead{$\chi^2$} & \colhead{$\hat{R}$} & \colhead{SLSN} & \colhead{SN~II} & \colhead{SN~IIn} & \colhead{SN~Ia} & \colhead{SN~Ibc}}
\startdata
PS0910012 & 0.0790 & 0.0073 & SN~Iax & SN~II & 0.830 & 4.8 & 3.0 & 0.000 & 0.830 & 0.002 & 0.056 & 0.112 \\
PSc010411 & 0.0740 & 0.0091 & FELT & SN~II & 0.515 & 6.2 & 1.0 & 0.000 & 0.515 & 0.000 & 0.097 & 0.388 \\
PSc040777 & 0.1680 & 0.0134 & TDE & SN~IIn & 0.772 & 1.2 & 1.0 & 0.000 & 0.036 & 0.772 & 0.186 & 0.006 \\
PSc080333 & 1.3883 & 0.0537 & Lensed SN~Ia & SLSN & 0.853 & 1.0 & 1.0 & 0.853 & 0.000 & 0.038 & 0.109 & 0.000 \\
PSc091902 & 0.1120 & 0.0563 & FELT & SN~Ibc & 0.477 & 1.4 & 1.0 & 0.000 & 0.322 & 0.003 & 0.198 & 0.477 \\
PSc120170 & 0.4046 & 0.0303 & TDE & SN~IIn & 0.637 & 2.4 & 1.0 & 0.023 & 0.015 & 0.637 & 0.323 & 0.002 \\
PSc150020 & 0.3230 & 0.0191 & FELT & SN~Ia & 0.741 & 1.5 & 1.0 & 0.046 & 0.000 & 0.140 & 0.741 & 0.073 \\
PSc340012 & 0.6460 & 0.0302 & FELT & SN~Ia & 0.704 & 1.3 & 1.0 & 0.070 & 0.001 & 0.186 & 0.704 & 0.039 \\
PSc350224 & 0.1010 & 0.0300 & FELT & SN~Ibc & 0.575 & 2.3 & 1.1 & 0.000 & 0.394 & 0.000 & 0.031 & 0.575 \\
PSc350352 & 0.4050 & 0.0118 & FELT & SN~Ia & 0.727 & 1.4 & 1.1 & 0.032 & 0.000 & 0.164 & 0.727 & 0.077 \\
PSc370290 & 0.0535 & 0.0310 & SN~Ibn & SN~Ia & 0.567 & 6.0 & 2.2 & 0.000 & 0.060 & 0.159 & 0.567 & 0.214 \\
PSc370330 & 0.1760 & 0.0193 & SN~IIb? & SN~Ibc & 0.779 & 0.9 & 1.0 & 0.000 & 0.046 & 0.037 & 0.138 & 0.779 \\
PSc440088 & 0.2750 & 0.0970 & FELT & SN~Ia & 0.692 & 1.8 & 1.0 & 0.003 & 0.022 & 0.066 & 0.692 & 0.217 \\
PSc570006 & 0.2693 & 0.0638 & FELT & SN~Ia & 0.682 & 2.3 & 1.1 & 0.027 & 0.013 & 0.173 & 0.682 & 0.105 \\
PSc570060 & 0.2450 & 0.0620 & FELT & SN~II & 0.813 & 3.1 & 1.0 & 0.000 & 0.813 & 0.012 & 0.055 & 0.120 \\
PSc580304 & 0.2960 & 0.0292 & FELT & SN~II & 0.400 & 3.5 & 1.0 & 0.000 & 0.400 & 0.089 & 0.385 & 0.126
\enddata
\end{deluxetable*}

\subsection{Rare Classes of Transients}\label{sec:rare}
Sixteen of our spectroscopically classified transients did not belong to any of the five classes we consider here: one SN~Iax \citep{narayan_displaying_2011}, two tidal disruption events (TDEs; \citealt{gezari_ultravioletoptical_2012,chornock_ultraviolet-bright_2014}), one lensed SN~Ia \citep{quimby_extraordinary_2013}, one SN~Ibn \citep{sanders_ps1-12sk_2013}, one possible SN~IIb \citepalias{villar_superraenn_2020},\footnote{Because SN~IIb is a time-dependent classification (i.e., hydrogen features weaken during the evolution of the SN) and most of our classifications are determined by a single spectrum per transient, it is likely that other SNe~IIb are ``misclassified'' as SNe~II or SNe~Ibc.} and ten fast-evolving luminous transients (FELTs; \citealt{drout_rapidly_2014}). Presumably there are additional examples of these types of transients in our test set, but because our classifier has no ability to identify them, they will contaminate our five photometric samples at a low level. By passing the light curves of known rare transients to the classifier, we can investigate how it might classify unknown rare transients.

Table~\ref{tab:other} lists our classification results for these \Nrare{} transients. The SN~Iax is classified as an SN~II with high confidence, likely because of its low luminosity. Both TDEs are classified as SNe~IIn, likely because of their high luminosity and slow evolution. The lensed SN~Ia is classified as an SLSN with high confidence, likely because of its high luminosity. The SN~Ibn is classified as an SN~Ia with relatively low confidence, likely because it peaks at about the same luminosity as SNe~Ia. The possible SN~IIb is classified as an SN~Ibc. The FELTs fall into all three classes that do not typically have slow evolution: SNe~Ia, Ibc, and II. FELTs span a range of peak luminosities ($-16.5 > M > -20\ \mathrm{mag}$; \citealt{drout_rapidly_2014}), and their fast evolution means that their light curves are not well sampled at the cadence of PS1-MDS. In general, we conclude that our classifier behaves as expected for these rare transients based their peak luminosities and evolution time scales, but we note that in these cases a high confidence does not imply a correct classification.

\subsection{Active Galactic Nuclei}
In constructing the set of ``SN-like'' transients, we excluded light curves with a history of variability, with the intention of removing active galactic nuclei (AGNs) from the test set. However, if any AGNs survived this qualitative cut, they would likely be classified as SLSNe due to their high luminosities and slow evolution. To check for this possibility, we inspect the host-galaxy spectra of the photometric SLSNe to look for broad emission lines (a signature of accretion onto the central supermassive black hole). Not all of the spectra have a high enough signal-to-noise ratio to identify broad lines, but in at least 17 cases, they are visible. Of these, 14 of the transients are within $1''$ of the host center---PSc000478, PSc010120, PSc010186, PSc020026, PSc030013, PSc052281, PSc110163, PSc130394, PSc130732, PSc350614, PSc390545, PSc400050, PSc480585, PSc550061 (the latter is shown in Figure~\ref{fig:lcs})---meaning that the AGN and the transient may be one and the same. Many (but not all) of these light curves are near the detection threshold, which could either indicate a nuclear SN (or even a TDE) that is faint compared to its AGN host, or a slight increase in the luminosity of the AGN itself. Because we cannot distinguish between these two cases, we urge caution in using these classification results.

\section{Conclusions}\label{sec:conclusion}
We have presented the SN photometric classification package Superphot, based on the algorithm of \citetalias{villar_supernova_2019}. Training and then validating the classifier on \Ntrain{} spectroscopically classified SNe from the Pan-STARRS1 PS1-MDS, we find that it has an overall accuracy of \accuracy{} and completenesses (purities) of \completenessII{} (\purityII{}) for SNe~II, \completenessIa{} (\purityIa{}) for SNe~Ia, \completenessSLSN{} (\puritySLSN{}) for SLSNe, \completenessIIn{} (\purityIIn{}) for SNe~IIn, and \completenessIbc{} (\purityIbc{}) for SNe~Ibc. We then apply this to \Ntest{} previously unclassified transients from PS1-MDS for which we have robust host-galaxy redshifts, resulting in \photIa{} photometrically classified SNe~Ia, \photII{} SNe~II, \photIbc{} SNe~Ibc, \photIIn{} SNe~IIn, and \photSLSN{} SLSNe.

In the process of validating our results, we raised several issues that will be relevant to future photometric classification efforts.
\begin{enumerate}
    \item A small misclassification rate of SNe~Ia can easily dominate photometric samples of minority classes like SNe~Ibc.
    \item SNe~II overlap significantly with SNe~Ibc in feature space, likely due to the subset of SNe~II with linearly declining light curves.
    \item Adopting a threshold on the classification confidence can improve the completeness and purity of the photometric samples, but an analogous threshold on the number of light-curve points is not as effective.
    \item Agreement between two classifiers on a given transient is not necessarily an indication that they are correct; they may both be biased to misclassify certain transients in the same way.
    \item Users should take into account the relative rates of different classes of SNe in addition to the photometric classification probabilities.
    \item Transients belonging to none of the target classes can be misclassified into one of these classes with relatively high confidence.
    \item AGNs may be a significant contaminant in photometrically classified SLSN samples.
\end{enumerate}

Along with \citetalias{villar_superraenn_2020}, this is the first application of a multiclass machine-learning classifier to a large photometric data set. As such, it serves as an example of the utility (and also the challenges) of photometric classification in the era of large time-domain surveys. Given that currently only a small fraction of transients discovered are classified spectroscopically, and the reality that this fraction will only decrease as discovery rates increase, we will have to increasingly rely on methods like this to extract as much science as possible from our data.

In addition, the photometric samples presented here are among the largest in the literature for each class, demonstrating the power of photometric classification to enable statistical studies of SNe. Importantly, however, each classification comes with an uncertainty. In the coming years, our field will have to learn how to handle exactly this type of photometric data set, when we will never know with certainty whether an individual classification is ``correct.'' No single classifier will likely outperform the others for all use cases, but continued testing of algorithms individually and in combination will demonstrate how best to apply them toward a specific science goal.

\facilities{ADS, NED, PS1}
\defcitealias{astropy_collaboration_astropy_2018}{Astropy Collaboration 2018}
\defcitealias{theano_development_team_theano_2016}{Theano Development Team 2016}
\software{ArviZ \citep{kumar_arviz_2019}, Astropy \citepalias{astropy_collaboration_astropy_2018}, extinction \citep{barbary_extinction_2016}, imbalanced-learn \citep{lemaitre_imbalanced-learn_2017}, IPython \citep{perez_ipython_2007}, Matplotlib \citep{hunter_matplotlib:_2007}, NumPy \citep{oliphant_guide_2006}, PyMC3 \citep{salvatier_probabilistic_2016}, RVSAO \citep{kurtz_rvsao_1998}, scikit-learn \citep{pedregosa_scikit-learn_2011}, SciPy \citep{virtanen_scipy_2020}, Theano \citepalias{theano_development_team_theano_2016}, \texttt{tqdm} \citep{da_costa-luis_tqdm_2019}}

\acknowledgments
We thank Jessica Mink and Brian Hsu for assisting with the host-galaxy redshifts. We also thank the authors of the ``Scientific Python Cookiecutter'' tutorial for advice on how to document, package, and release the Superphot package. The Berger Time-Domain Group is supported in part by NSF grant AST-1714498 and NASA grant NNX15AE50G. We acknowledge partial funding support from the Harvard Data Science Initiative. G.H. thanks the LSSTC Data Science Fellowship Program, which is funded by LSSTC, NSF Cybertraining grant \#1829740, the Brinson Foundation, and the Moore Foundation; his participation in the program has benefited this work. F.D.\ thanks the SAO REU program, funded in part by the National Science Foundation REU and Department of Defense ASSURE programs under NSF grant No.\ AST-1852268 and by the Smithsonian Institution. V.A.V.\ acknowledges support by the Ford Foundation through a Dissertation Fellowship and the Simons Foundation through a Simons Junior Fellowship (\#718240). D.O.J.\ is supported by a Gordon and Betty Moore Foundation postdoctoral fellowship at the University of California, Santa Cruz. The UCSC team is supported in part by NASA grants 14-WPS14-0048, NNG16PJ34C, NNG17PX03C; NSF grants AST-1518052 and AST-1815935; NASA through grant No.\ AR-14296 from the Space Telescope Science Institute, which is operated by AURA, Inc., under NASA contract NAS 5-26555; the Gordon \& Betty Moore Foundation; the Heising-Simons Foundation; and by fellowships from the Alfred P.\ Sloan Foundation and the David and Lucile Packard Foundation to R.J.F.  R.L.\ is supported by a Marie Sk\l{}odowska-Curie Individual Fellowship within the Horizon 2020 European Union (EU) Framework Programme for Research and Innovation (H2020-MSCA-IF-2017-794467). D.M. acknowledges NSF support from from grants PHY-1914448 and AST-2037297.

The Pan-STARRS1 Surveys (PS1) and the PS1 public science archive have been made possible through contributions by the Institute for Astronomy, the University of Hawaii, the Pan-STARRS Project Office, the Max-Planck Society and its participating institutes, the Max Planck Institute for Astronomy, Heidelberg, and the Max Planck Institute for Extraterrestrial Physics, Garching, Johns Hopkins University, Durham University, the University of Edinburgh, Queen's University Belfast, the Center for Astrophysics \textbar{} Harvard \& Smithsonian, Las Cumbres Observatory, the National Central University of Taiwan, the Space Telescope Science Institute, the National Aeronautics and Space Administration under grant No.\ NNX08AR22G issued through the Planetary Science Division of the NASA Science Mission Directorate, the National Science Foundation grant No.\ AST-1238877, the University of Maryland, E\"otv\"os Lor\'and University (ELTE), Los Alamos National Laboratory, and the Gordon and Betty Moore Foundation.

\appendix
\section{Glossary}\label{sec:glossary}
For each SN, the classifier gives a set of five \textbf{classification probabilities} ($p_\Phi$) corresponding to each of the five photometric classes. They sum to 1 and are listed in Tables~\ref{tab:results}--\ref{tab:other}:
$$p_\Phi \in \{p_\mathrm{SLSN}, p_\mathrm{SNII}, p_\mathrm{SNIIn}, p_\mathrm{SNIa}, p_\mathrm{SNIbc}\}; \quad \sum_\Phi p_\Phi = 1.$$
An SN's photometric classification ($\Phi$) is determined by its highest classification probability. We take this probability to be the classification \textbf{confidence} ($p$):
$$p \equiv \max\{p_\Phi\}.$$

$N_{S\Phi}$ is the number of SNe with spectroscopic classification $S$ and photometric classification $\Phi$; these are the elements of the \textbf{confusion matrix} (the integers in Figure~\ref{fig:confusion}). $N_S \equiv \sum_\Phi N_{S\Phi}$ is the total number of SNe with spectroscopic classification $S$, $N_\Phi \equiv \sum_S N_{S\Phi}$ is the total number of SNe with photometric classification $\Phi$, and $N \equiv \sum_S \sum_\Phi N_{S\Phi}$ is the total sample size. $N_{S=\Phi}$ is the number of correctly classified SNe in a given class.

We discuss four performance metrics for our classifier, all of which range from 0 to 1:
\begin{enumerate}
    \item Completeness ($C_S$) is the fraction of a given spectroscopic class that appears in the equivalent photometric class:
    $$C_S \equiv \frac{N_{S=\Phi}}{N_S}.$$
    \item Purity ($P_\Phi$) is the fraction of a given photometric class that belongs to the equivalent spectroscopic class:
    $$P_\Phi \equiv \frac{N_{S=\Phi}}{N_\Phi}.$$
    \item Accuracy ($A$) is the total fraction of correctly classified SNe:
    $$A \equiv \frac{\sum_{S=\Phi} N_{S\Phi}}{N}.$$
    A variant of accuracy is agreement ($A'$), in which we compare two photometric classifications to each other, rather than comparing a photometric classification to a spectroscopic classification.
    \item The (macro-averaged) $F_1$ score is the average of the harmonic means of the completeness and purity of each class:
    $$F_1 \equiv \frac{1}{5} \sum_{S=\Phi} \frac{2}{C_S^{-1} + P_\Phi^{-1}} = \frac{1}{5} \sum_{S=\Phi} \frac{2N_{S\Phi}}{N_S + N_\Phi}.$$
\end{enumerate}

\section{Agreement between Two Classifiers}\label{sec:agree}
In Section~\ref{sec:compare}, we compared our photometric classifications to those of \citetalias{villar_superraenn_2020}. Here we derive the expectation for such a comparison given the confusion matrices for each classifier calculated from cross-validation.

The version of the confusion matrix with completeness on the diagonal (Figure~\ref{fig:confusion}, top) shows the probability $p(\Phi|S)$ of our classifier giving the photometric classification $\Phi$ for an SN with spectroscopic classification $S$. In this section, we will refer to this as the ``completeness matrix'' $\bm{C}$. The version of the confusion matrix with purity on the diagonal (Figure~\ref{fig:confusion}, center) shows the probability $p(S|\Phi)$ of an SN having spectroscopic classification $S$ if we gave it a photometric classification $\Phi$. In this section, we will refer to this as the ``purity matrix'' $\bm{P}$.

The ``agreement matrix'' $\bm{A}$ we wish to derive will show the probability $p(\Phi'|\Phi)$ of another classifier giving photometric classification $\Phi'$ to a transient that we classify as $\Phi$. Using the chain rule of probability, we can write
$$p(\Phi'|\Phi) = \frac{p(\Phi,\Phi')}{p(\Phi)}.$$
The joint probability $p(\Phi, \Phi')$ cannot be separated because the classifiers are not independent. (If they were, they would not be good classifiers.) However, we assume that the classifiers are independent for a given spectroscopic class, meaning that, if they are biased, they are not biased in the same way. We can then obtain a separable joint probability by undoing the marginalization over spectroscopic classification:
$$p(\Phi'|\Phi) = \sum_S\frac{p(\Phi,\Phi'|S)p(S)}{p(\Phi)} = \sum_S\frac{p(\Phi|S)p(\Phi'|S)p(S)}{p(\Phi)}.$$
Lastly, we simplify using Bayes's theorem \citep{bayes_essay_1763}:
$$p(\Phi'|\Phi) = \sum_S p(S|\Phi)p(\Phi'|S),$$
or in matrix notation,
$$\bm{A} = \bm{P}^\mathrm{T} \bm{C}',$$
where $\bm{C}'$ is the completeness matrix of the other classifier. Note that this matrix depends on the breakdown of spectroscopic classes in the data sets used to validate these classifiers. Figure~\ref{fig:agree} (top left) shows the expected agreement matrix between our classifier and that of \citetalias{villar_superraenn_2020}, assuming that our test set has similar class fractions to the training set.

\section{Feature Importances}\label{sec:importance}
Of interest to developers of future photometric classification algorithms is the question of which features are most powerful for classifying SNe. \citetalias{villar_supernova_2019} explore several combinations of features, and we adopt their best-performing classifier, which uses peak absolute magnitude in the $griz$ filters plus coefficients of the top six principal components of the $griz$ light curves. Here we explore the relative importances of each of those features.

\begin{figure}
    \centering
    \includegraphics[width=\columnwidth]{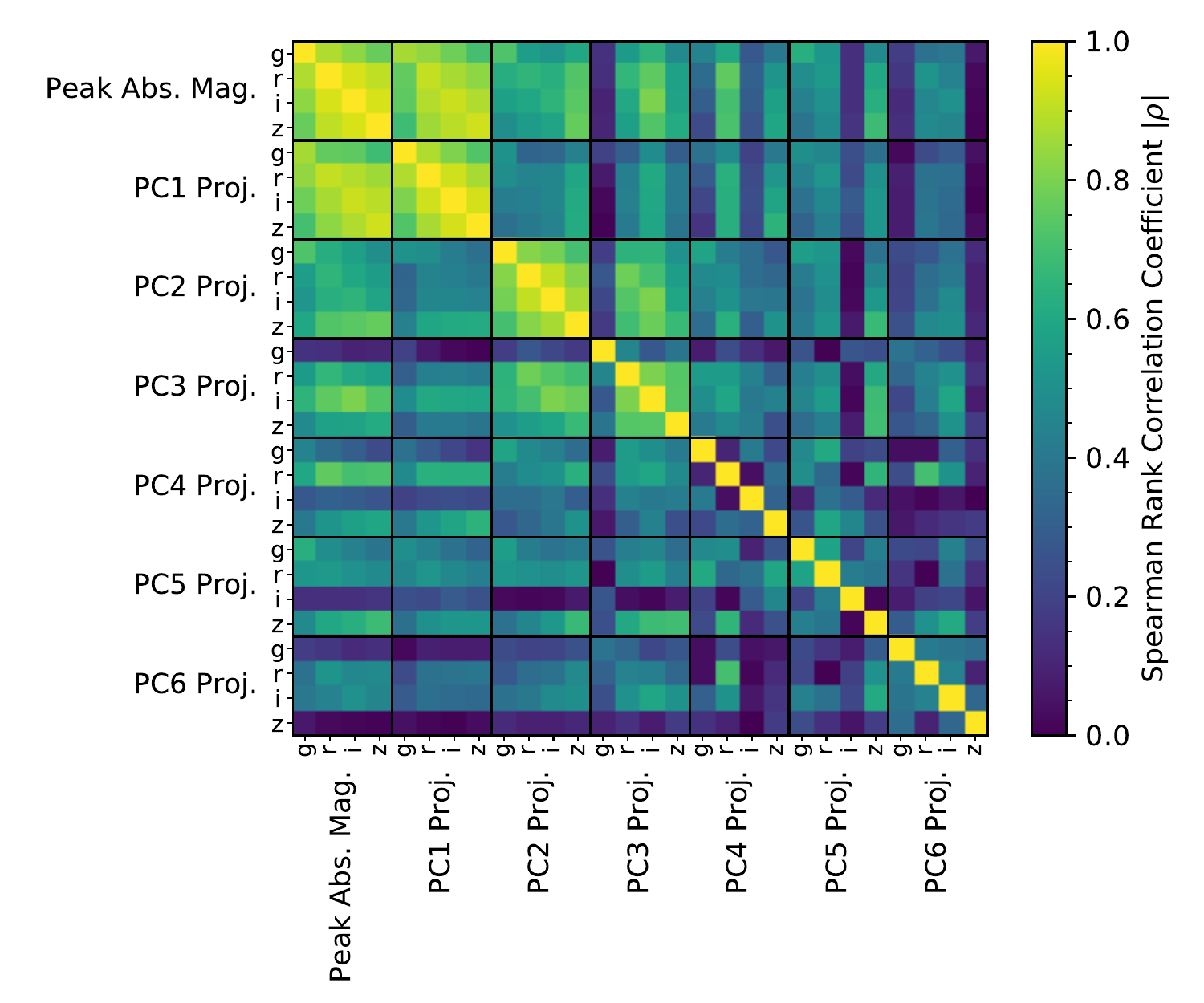}
    \includegraphics[width=\columnwidth]{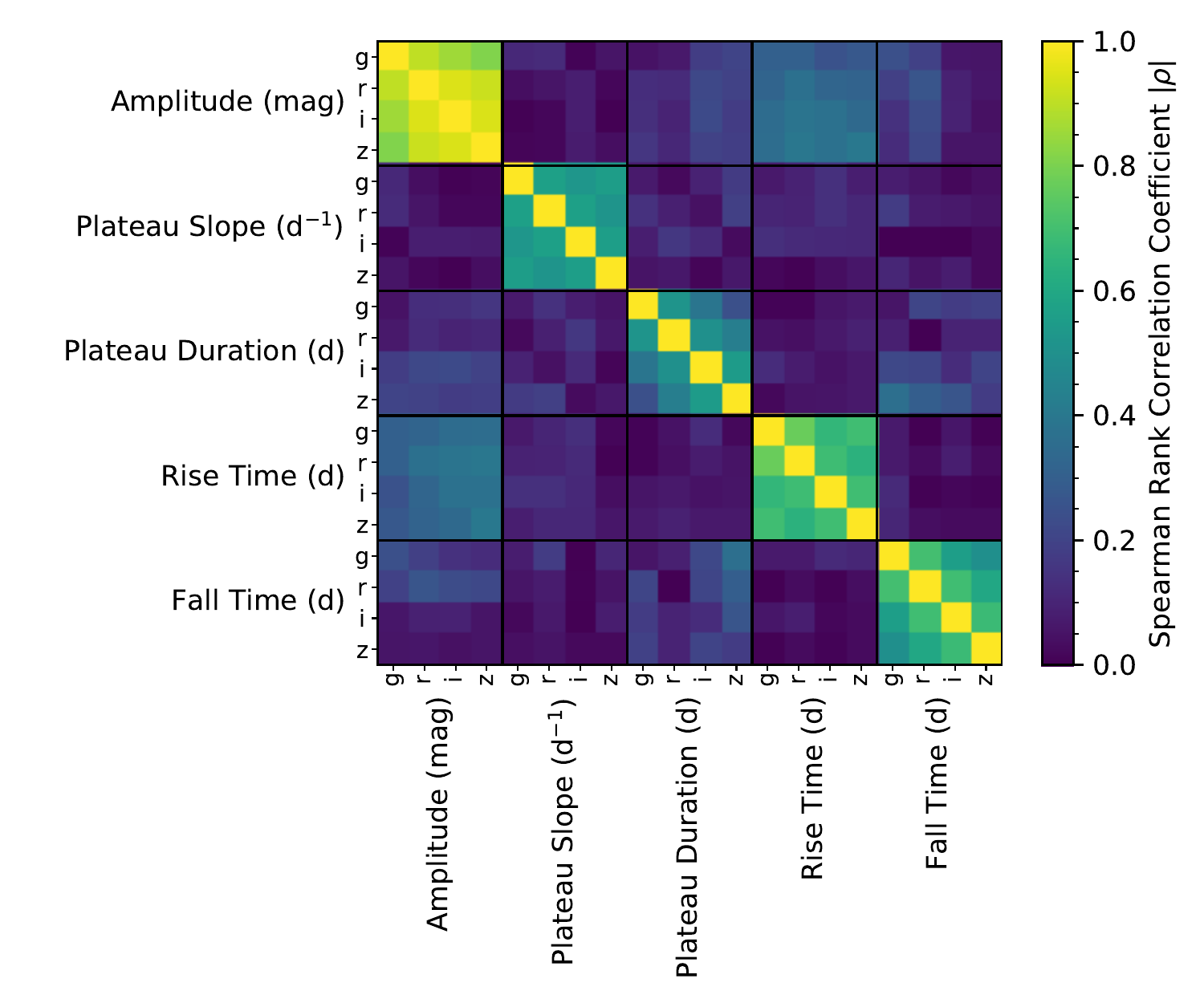}
    \caption{Top: absolute values of the Spearman rank correlation coefficients between each pair of features in our training set. Note the strong correlations between the four filters, as well as a correlation between the peak absolute magnitude and the projection onto the first principal component of the light curves. Correlations among the remaining features are weak. Bottom: the equivalent correlation matrix for the model parameters.}
    \label{fig:correlation}
\end{figure}

The importance of a feature cannot be defined independently of the other features used. For example, if two features are perfectly correlated, a classifier may arbitrarily consider one of them to be very important and the other to be useless. Therefore, before calculating feature importances, we must examine correlations between the features in our set. Figure~\ref{fig:correlation} shows the absolute value of the \cite{spearman_proof_1904} rank correlation coefficient between every pair of features (top) and model parameters (bottom) in our training set.

As we might expect, there are strong correlations between the four filters, both because physics demands a relatively smooth SED and because our two-iteration fitting method forces the light-curve models in the four filters to be more similar. This means it is not possible to judge whether, for example, the $g$ peak absolute magnitude is more important than the $r$ peak absolute magnitude. In addition, we find that the peak absolute magnitudes are strongly correlated with the light curves' projection onto their principal component. This tells us that most of the variation among the light curves can be attributed to differences in overall luminosity, rather than differences in shape.

\begin{figure}
    \centering
    \includegraphics[width=\columnwidth]{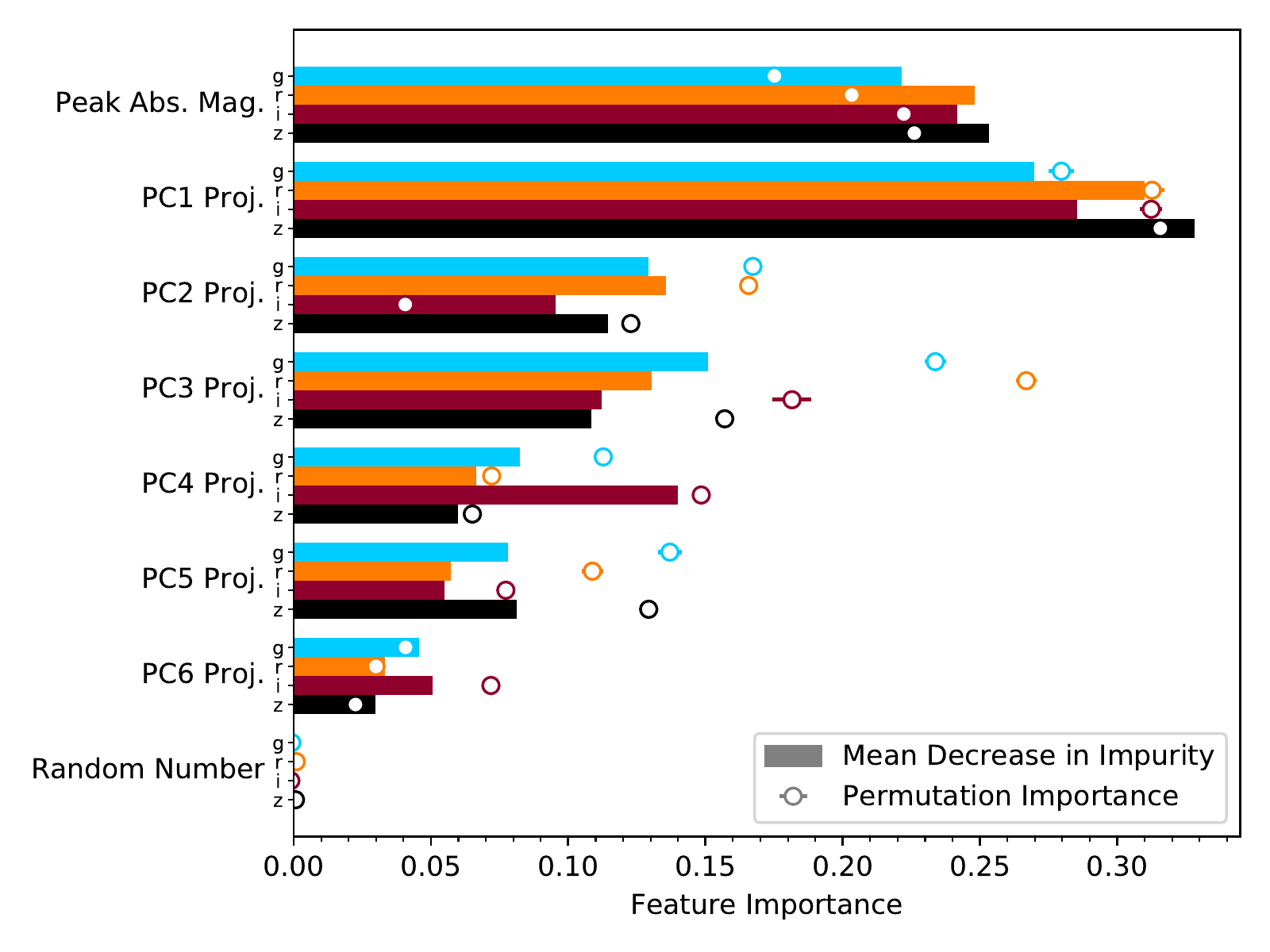}
    \caption{Feature importances for our training set. Cross-filter comparisons, as well as comparison between the peak absolute magnitude and the projection onto the first principal component, are meaningless due to the correlations at left. Nonetheless, it is clear that peak absolute magnitude and the projection onto the first principal component are the most important in all filters.}
    \label{fig:importance}
\end{figure}

To obtain meaningful feature importances, we retrained our classifier on only one filter (seven features) at a time. We then calculated two measures of feature importance for each of the seven features: mean decrease in impurity \citep{louppe_understanding_2015} and permutation importance \citep{breiman_random_2001}. We also calculate the permutation importance of a random feature, for comparison, which is consistent with zero for all filters. The results are shown in Figure~\ref{fig:importance}.

In all cases, the peak absolute magnitude and the projection onto the first principal component---we cannot compare these to each other because they are correlated---are by far the most discriminating between the classes. The remaining principal component coefficients contribute roughly in order of their rank. We again emphasize that cross-filter comparisons are meaningless in our analysis, but the relative importances of the features are similar for each filter.

\begin{figure*}
    \centering
    \includegraphics[width=\textwidth]{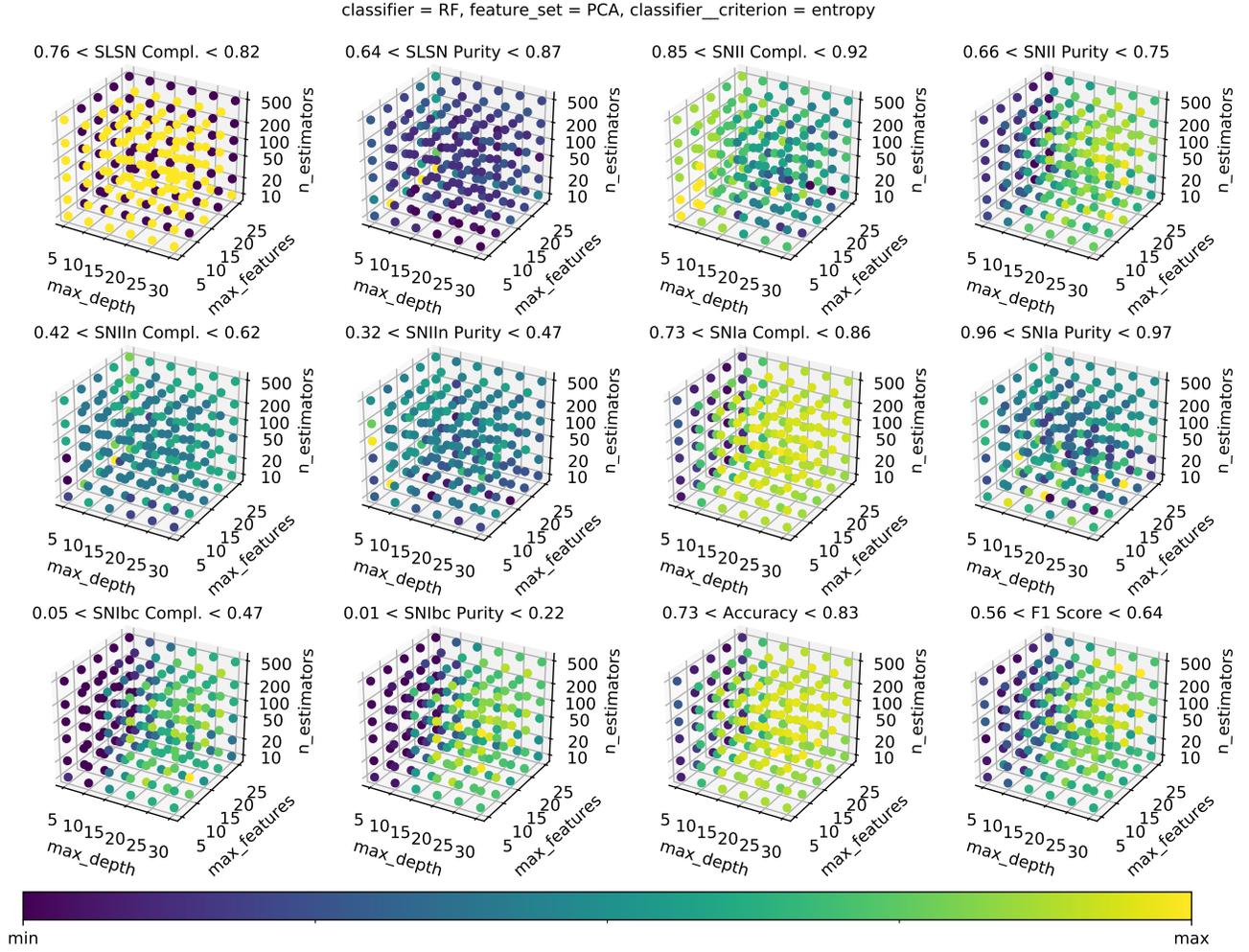}
    \caption{Results of varying the hyperparameters of our classifier over a three-dimensional grid, as measured by 12 metrics. The range of each metric is shown above each plot, and the color of each point corresponds to where it lies in that range. There is no single set of hyperparameters that optimizes all of the metrics, but in general, we find it important not to limit the maximum depth of the decision trees. (The data used to create this figure are available.)}
    \label{fig:hyperparameters}
\end{figure*}

\section{Hyperparameter Optimization}\label{sec:hyperparam}
Our random forest classifier has several hyperparameters that can be adjusted to obtain better results. We repeated our analysis, apart from the final classification, over a grid of four of these parameters (represented by their variable names in scikit-learn; \citealt{pedregosa_scikit-learn_2011}):
\begin{enumerate}
    \item $\texttt{criterion} \in \{\textrm{Gini impurity}, \textrm{entropy}\}$, the function to measure the quality of a split;
    \item $\mathtt{max\_depth} \in \{5, 10, 15, 20, 25, 29\}$, the maximum depth of a decision tree;
    \item $\mathtt{max\_features} \in \{5, 10, 15, 20, 25\}$, the number of features to consider when looking for the best split; and
    \item $\mathtt{n\_estimators} \in \{10, 20, 50, 100, 200, 500\}$, the number of decision trees in the random forest.
\end{enumerate}

\begin{deluxetable*}{ccCCCCCCCCCCCC}
\tablecaption{Other Classifiers and Feature Sets\label{tab:hyperparameters}}
\tablehead{\colhead{} & \colhead{} & \multicolumn{5}{c}{Completeness} & \multicolumn{5}{c}{Purity} & \colhead{} & \colhead{} \\[-13pt]
\colhead{Classifier} & \colhead{Feature Set} & \multicolumn{5}{c}{------------------------------------------------------------} & \multicolumn{5}{c}{------------------------------------------------------------} & \colhead{Accuracy} & \colhead{$F_1$ Score} \\[-12pt]
\colhead{} & \colhead{} & \colhead{SLSN} & \colhead{SN~II} & \colhead{SN~IIn} & \colhead{SN~Ia} & \colhead{SN~Ibc} & \colhead{SLSN} & \colhead{SN~II} & \colhead{SN~IIn} & \colhead{SN~Ia} & \colhead{SN~Ibc} & \colhead{} & \colhead{}}
\startdata
RF & PCA & \phantom{+}0.824 & \phantom{+}0.892 & \phantom{+}0.500 & \phantom{+}0.849 & \phantom{+}0.368 & \phantom{+}0.667 & \phantom{+}0.728 & \phantom{+}0.387 & \phantom{+}0.961 & \phantom{+}0.206 & \phantom{+}0.824 & \phantom{+}0.628 \\
RF & parameters & -0.176 & -0.075 & +0.125 & -0.010 & +0.105 & +0.067 & +0.024 & -0.133 & +0.002 & +0.094 & -0.016 & -0.009 \\
SVM & PCA & +0.118 & +0.011 & -0.125 & +0.012 & -0.053 & -0.326 & +0.050 & +0.022 & -0.002 & +0.147 & +0.007 & -0.035 \\
MLP & PCA & -0.176 & -0.043 & -0.083 & -0.045 & +0.053 & -0.020 & -0.016 & -0.143 & +0.012 & -0.058 & -0.047 & -0.062
\enddata
\end{deluxetable*}

For each of these 360 classifiers, we calculated 12 different metrics on our training set: completeness and purity in each of our 5 classes, accuracy, and $F_1$ score (see Appendix~\ref{sec:glossary} for definitions). Figure~\ref{fig:hyperparameters} shows the results when using entropy as the split criterion; switching to the Gini impurity was neutral or slightly worse in most cases. There is no clear winner in all 12 metrics, but in general, we find it important not to limit the depth of the decision trees. The final classifier used in our analysis uses an entropy criterion, no maximum depth, a maximum of 5 features, and 100 estimators.

In addition to optimizing the random forest (RF) hyperparameters, we repeated our analysis using two other classification algorithms: a support vector machine (SVM; with regularization parameter $C=1000$ and kernel coefficient $\gamma=0.1$) and a multilayer perceptron (MLP; with two hidden layers of 10 and 5 neurons and an L2 penalty parameter $\alpha=10^{-5}$). We also repeated our analysis using the model parameters directly as features, rather than performing a PCA. Table~\ref{tab:hyperparameters} lists the same 12 metrics for our baseline procedure (which has the highest $F_1$ score) in the top row and how these metrics change using the other classifiers and feature set. The RF with the model parameters and the SVM perform similarly to our baseline algorithm, but with some important weaknesses (e.g., low SLSN completeness and purity, respectively). The MLP performs worse in most metrics.

\bibliography{My_Library,galaxy-surveys}

\end{document}